\documentclass[superscriptaddress, twocolumn, nofootinbib]{revtex4-2}

\usepackage{amsmath}
\usepackage{bm}
\usepackage{booktabs}
\usepackage{braket}
\usepackage{cancel}
\usepackage{color}
\usepackage{float}
\usepackage[colorlinks=true, allcolors=blue]{hyperref}
\usepackage{cleveref}
\usepackage{graphics}

\newcommand{\fdd}[1][]{\left(-\frac{\partial f_{0}}{\partial\mathcal{E}_{\bm{k}#1}}\right)}

\begin{document}

\title{On non-local electrical transport in anisotropic metals}
\author{Graham Baker}
\email{graham.baker@cpfs.mpg.de}
\affiliation{Max Planck Institute for Chemical Physics of Solids, Dresden, Germany}
\author{Davide Valentinis}
\affiliation{
    Institute for Theory of Condensed Matter
    \& Institute for Quantum Materials and Technologies, Karlsruhe Institute of Technology, Karlsruhe, Germany}
\author{Andrew P. Mackenzie}
\affiliation{Max Planck Institute for Chemical Physics of Solids, Dresden, Germany}
\affiliation{Scottish Universities Physics Alliance, School of Physics and Astronomy, University of St Andrews, United Kingdom}

\begin{abstract}
    We discuss various aspects of non-local electrical transport in anisotropic metals. For a metal with circular Fermi surface, the scattering rates entering the local conductivity and viscosity tensors are well-defined, corresponding to eigenfrequencies of the linearized collision operator. For anisotropic metals, we provide generalized formulas for these scattering rates and use a variational approximation to show how they relate to microscopic transition probabilities. We develop a simple model of a collision operator for a metal of arbitrary Fermi surface with finite number of quasi-conserved quantities, and derive expressions for the wavevector-dependent conductivity $\sigma(q)$ and the spatially-varying conductivity $\sigma(x)$ for a long, narrow channel. We apply this to the case of different rates for momentum-conserving and momentum-relaxing scattering, deriving closed-form expressions for $\sigma(q)$ and $\sigma(x)$---beyond generalizing from circular to arbitrary Fermi surface geometry, this represents an improvement over existing methods which solve the relevant differential equation numerically rather than in closed form. For the specific case of a diamond Fermi surface, we show that, if transport signatures were interpreted via a model for a circular Fermi surface, the diagnosis of the underlying transport regime would differ based on experimental orientation and based on whether $\sigma(q)$ or $\sigma(x)$ was considered. Finally, we discuss the bulk conductivity. While the common lore is that ``momentum''-conserving scattering does not affect bulk resistivity, we show that \textit{crystal momentum}-conserving scattering---such as normal electron-electron scattering---can affect the bulk resistivity for an anisotropic Fermi surface. We derive a simple formula for this contribution. 
\end{abstract}

\maketitle

\section{Introduction}

As is common with outstandingly far-sighted science, the pioneering papers of Gurzhi \cite{Gurzhi1963,Gurzhi1968} on the possibility of viscous electronic transport in ultra-high purity metals were far ahead of their time. When he wrote them, there were few, if any, suitable material platforms on which to test his ideas. The first to arrive, three decades later, were the high purity semiconductor two-dimensional electron gases (2DEGs) on which Molenkamp and de Jong performed their intriguing experiments using current heating to raise the electron temperature and reach the viscous regime \cite{Molenkamp1994,DeJong1995}. Over the past decade, there have been rapid developments in the study of other materials with extremely low impurity scattering rates, such as graphene \cite{Crossno2016,Bandurin2016,KrishnaKumar2017,Sulpizio2019,Ku2020,Kumar2022}, delafossites such as PdCoO$_2$ and PtCoO$_2$ \cite{Moll2016,Bachmann2019,McGuinness2021,Bachmann2022,Baker2022,Zhakina2023}, and semimetals such as WP$_2$ and WTe$_2$ \cite{Ali2014,Zhu2015,Gooth2017,VanDelft2021,Vool2021}. Many intriguing signatures of non-local transport have been observed, and perhaps the biggest lesson learned through the process is that two famous non-local regimes, the `Gurzhi', `viscous', or `Poiseuille' regime and the `ballistic' or `Knudsen' regime, are not nearly as distinct as had previously been assumed. Since the viscous regime is the newer and more exotic, a common path has been for a signature claimed to be an unambiguous proof of viscous behavior to be subsequently realized to be either primarily ballistic in origin or at least to be explicable by ballistic physics in combination with other previously ignored real-world complications.

One aspect of several of the new materials whose importance has only been fully appreciated in the past few years is Fermi surface anisotropy, which is particularly relevant to the study of PdCoO$_2$, PtCoO$_2$, WP$_2$, and WTe$_2$. Indeed, in the delafossites, a seemingly minor anisotropy in the Fermi surface geometry has large physical consequences \cite{Bachmann2019,McGuinness2021,Bachmann2022,Baker2022,Valentinis2023,Zhakina2023}. Analysis of transport properties using the assumption of isotropic Fermi surfaces has been shown to be inadequate in such situations, strongly motivating the construction of analysis models capable of taking Fermi surface anisotropy into account. Although considerable progress has been made in that regard \cite{Bachmann2019,Qi2021,Bachmann2022,Varnavides2022,Varnavides2022b,Baker2022,Valentinis2023}, it is desirable to find closed-form expressions for as many of the relevant quantities as possible, to increase the efficiency and transparency of the numerical calculations that must be performed. In this paper, we make two contributions to that process, with the aim of furthering Gurzhi's goal of obtaining a full understanding of non-local transport beyond the standard ohmic regime of metals.

\section{Crystal momentum and group velocity in anisotropic metals}

\begin{figure}
    \includegraphics{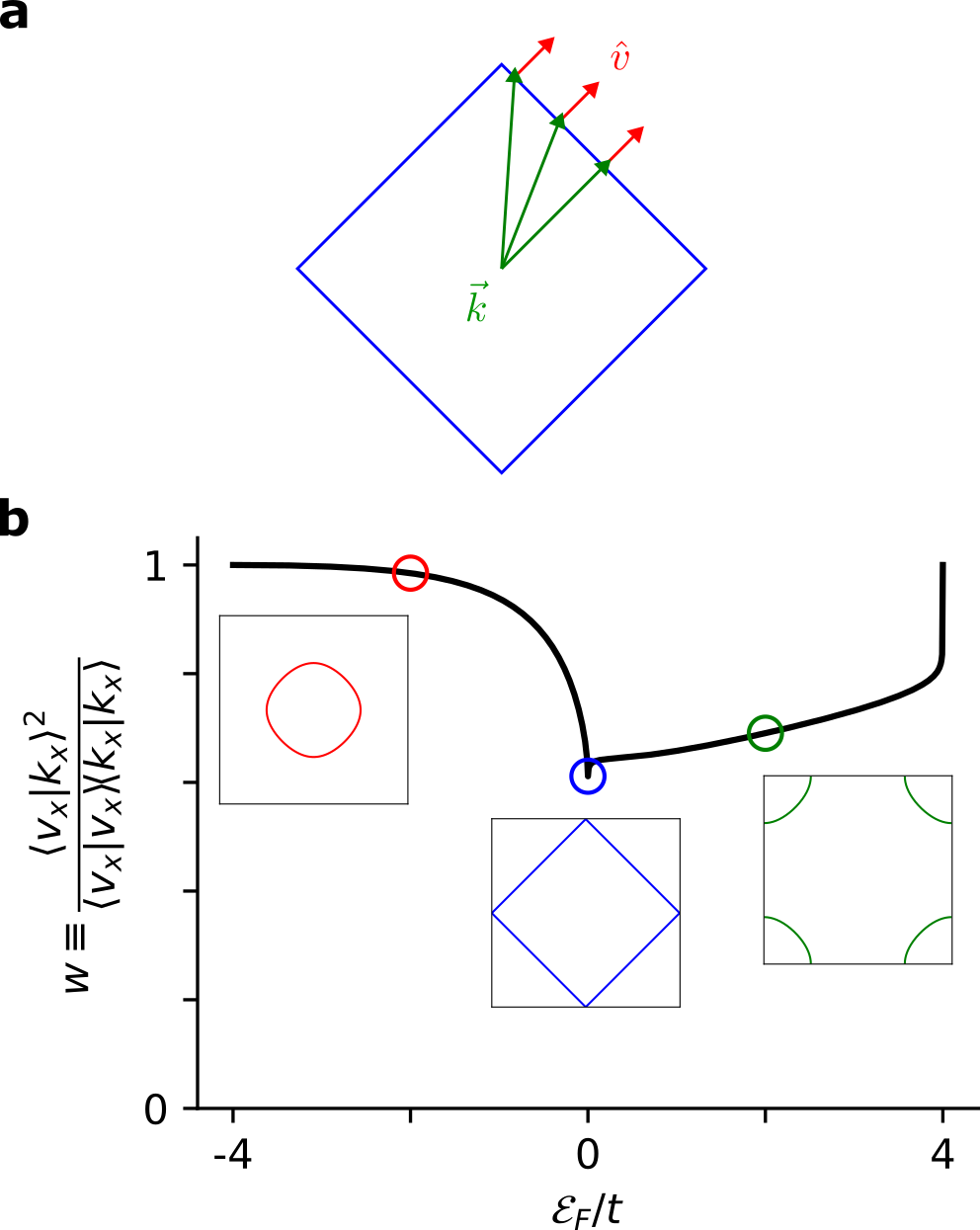}
    \caption{(a) Crystal momentum $\vec{k}$ and group velocity $\vec{v}$ are not necessarily parallel for an anisotropic Fermi surface, as illustrated here for a diamond Fermi surface. The unit group velocity vector $\hat{v}$ is always perpendicular to the Fermi surface. (b) The average overlap of crystal momentum and group velocity on the Fermi surface as a function of Fermi energy for a tight-binding model on a square lattice. The corresponding Fermi surface is shown for selected values of $\mathcal{E}_{F}/t$. While this band structure is known to be particle-hole symmetric, $w$ as defined here is not an even function of $\mathcal{E}_{F}/t$. This reflects a subtlety arising in anisotropic metals: while group velocity is uniquely defined, crystal momentum depends on the choice of primitive cell. Here we have used the common choice of taking the $\Gamma$ point as the origin when defining crystal momentum.}
    \label{fig:kv_overlap}
\end{figure}

In anisotropic metals, care is required to distinguish between several quantities. We consider Bloch electrons, obeying $\hat{H}\psi_{\bm{k}}=\mathcal{E}_{\bm{k}}\psi_{\bm{k}}$ where $\hat{H}$ is the single-particle Hamiltonian $\hat{H}=(-\hbar^{2}/2m)\nabla^{2}+V(\bm{r})$ and $V(\bm{r})$ is a lattice-periodic potential. Three related quantities are the crystal momentum $\bm{k}$, group velocity $\bm{v}_{\bm{k}}=(1/\hbar)\nabla_{\bm{k}}\mathcal{E}_{\bm{k}}$, and the momentum operator $\hat{\bm{p}}\equiv(\hbar/i)\nabla$. 
For a free electron metal with $\mathcal{E}_{\bm{k}}=(\hbar k)^{2}/2m$, these three vector quantities are parallel: $\bm{p}=\hbar\bm{k}=m\bm{v}$ (where the momentum $\bm{p}$ is the eigenvalue of $\hat{\bm{p}}$). 
For an anisotropic metal, $\bm{k}$ and $\bm{v}$ are not in general parallel, as illustrated in \cref{fig:kv_overlap}a, and the Bloch states are not eigenstates of the momentum operator $\hat{\bm{p}}$. However, it can be shown that the expectation value of the momentum operator is related to the group velocity: $\langle\hat{\bm{p}}\rangle_{\bm{k}}\equiv\int d^{3}\bm{r}\,\psi_{\bm{k}}^{*}(\bm{r})\,\hat{\bm{p}}\,\psi_{\bm{k}}(\bm{r})=m\bm{v}_{\bm{k}}$ \cite{Ashcroft1976}. 
Finally, electrical current, often the physically observable quantity, is given by the total group velocity of all electrons. Throughout this paper, we will explore how the difference between crystal momentum and group velocity leads to novel subtleties and phenomena in the transport properties of anisotropic metals.

As a measure of the similarity between crystal momentum and group velocity, we introduce a quantity $w$ which is a thermally-averaged overlap between the two quantities:
\begin{equation}
    w
    \equiv\frac{{\langle}v_{x}|k_{x}{\rangle}^{2}}{{\langle}v_{x}|v_{x}{\rangle\!{\langle}k_{x}|k_{x}{\rangle}}}
\end{equation}
where we have defined the inner product 
\begin{equation}
    {\langle}b|a{\rangle}
    \equiv\sum_{\bm{k}}\fdd b_{\bm{k}}^{*}a_{\bm{k}} .
\end{equation}
%
In the degenerate limit $T{\ll}T_{F}$, $(-\partial f_{0}/\partial\mathcal{E}_{\bm{k}})\to\delta(\mathcal{E}_{\bm{k}}-\mathcal{E}_{F})$ so that the average is restricted to the Fermi surface. 

For the purpose of illustration, in \cref{fig:kv_overlap}b we have evaluated $w$ in the degenerate limit as the Fermi surface geometry evolves as a function of Fermi energy $\mathcal{E}_{F}$ for a nearest-neighbor tight-binding model on a square lattice:
\begin{equation}\label{eq:tight_binding}
    \mathcal{E}_{\bm{k}}
    =\mathcal{E}_{F}-2t[\cos(k_{x}a)+\cos(k_{y}a)] .
\end{equation}
Aside from being a simple average measure of the degree to which crystal momentum and group velocity differ, later on, we see that $w$ also takes on a specific physical significance in certain contexts. However, as is evident in \cref{fig:kv_overlap} in which $w$ breaks the particle-hole symmetry of the band structure, $k$ and thus $w$ are not uniquely defined but rather depend on the choice of primitive cell. 

\section{Experimental quantities}\label{sec:exp_quants}

The fundamental quantity in non-local electrical transport is the non-local conductivity $\sigma(\bm{r}-\bm{r}')$ which enters the generalized version of Ohm's law:
\begin{equation}\label{eq:ohm_r}
    J_{i}(\bm{r})
    =\int_{\{\bm{r}_{0}\}} d^{d}x'\,
    \sigma_{ij}(\bm{r}-\bm{r}')E_{j}(\bm{r}') .
\end{equation}
The range of integration $\{\bm{r}_{0}\}$ depends on the geometry of the sample and the nature of electronic scattering at the sample's boundaries. If the range is taken to be from $-\infty$ to $\infty$, i.e. if translational invariance is assumed, \cref{eq:ohm_r} can be Fourier transformed to give
\begin{equation}
    J_{i}(\bm{q})
    =\sigma_{ij}(\bm{q})E_{j}(\bm{q}) .
\end{equation}
The wavevector-dependent conductivity can also be thought of as describing the response to a monochromatic electric field $E_{0}e^{i\bm{q}\cdot\bm{r}}$. 

While $\sigma(q)$ is often the more straightforward quantity to calculate, real samples break translational invariance. To connect with experiments, \cref{eq:ohm_r} should in principle be solved with appropriate boundary conditions coming from a treatment of electron-boundary scattering. However, in practice, such solutions have only been found for simple geometries. An approach taken by some authors \cite{Nazaryan2021,Qi2021} to describe complex geometries, e.g. electron flow through one or more slits, has been to calculate $\sigma(q)$ and to introduce a fictional electrical field to enforce approximate boundary conditions. 

On the other hand, there are two particularly simple measurement geometries for which a solution to the Boltzmann equation is possible using boundary conditions based on electron-boundary scattering.
Incidentally, these are the two geometries originally considered by \citet{Gurzhi1968}.
The first is the flow of DC electrical current down a long, narrow channel. Accounting for electron-boundary scattering at the two walls, one can calculate the conductivity $\sigma(x)$---the current across the channel normalized by the constant electric field---or its spatial average $\overline{\sigma(x)}$ as can be measured via resistivity.

The second experimental scenario for which treating electron-boundary scattering is possible is the surface impedance of a semi-infinite metal \cite{Baker2022,Valentinis2023}. Despite the broken symmetry due to the planar boundary of the medium, the surface impedance can nonetheless be expressed in terms of the wavevector-dependent conductivity \cite{Reuter1948,Dingle1953}. For a conductivity $\sigma(q)\sim q^{-\alpha}$, the surface impedance follows $Z\sim\omega^{\beta}\exp[-i(\pi/2)\beta]$ with $\beta=(1+\alpha)/(2+\alpha)$ and, to lowest order, only its prefactor depends on the nature of the boundary scattering (see \cref{sec:impedance} for a derivation of this scaling relation). 

Throughout this paper, we will focus on two quantities, motivated by the two above-mentioned experiments: the wavevector-dependent conductivity $\sigma(q)$ describing the response to a monochromatic electric field, and the conductivity $\sigma(x)$ of a finite-width channel. Surprisingly, we will find that the conclusions drawn about the nature of the transport regime from one quantity or the other do not always match for anisotropic metals.

\section{Boltzmann equation}

To calculate the electrical conductivities of anisotropic metals covering the ohmic, viscous, and ballistic regimes, we will solve the Boltzmann equation in conjunction with a phenomenological model of momentum-relaxing and momentum-conserving scattering. Here we introduce the concepts and notation required for following sections.

The Boltzmann equation describing the time evolution of the electronic distribution function $f_{\bm{k}}(\bm{r},t)$ under the influence of an electric field $\bm{E}$ is
\begin{equation}
    \partial_{t}f_{\bm{k}}
    +\bm{v}_{\bm{k}}\cdot\nabla_{\bm{r}}f_{\bm{k}}
    -\frac{e}{\hbar}\bm{E}\cdot\nabla_{\bm{k}}f_{\bm{k}}
    =-\mathcal{C}_{\bm{k}}[f_{\bm{k}}]
\end{equation}
where $\bm{k}$ is crystal momentum, $\bm{v}_{\bm{k}}=(1/\hbar)\nabla_{\bm{k}}\mathcal{E}_{\bm{k}}$ is group velocity, and $\mathcal{E}_{\bm{k}}$ is the electronic dispersion. The collision operator $\mathcal{C}_{\bm{k}}[f_{\bm{k}}]$ accounts for changes to $f_{\bm{k}}$ due to scattering.
We are interested in the linearized Boltzmann equation, which results from expanding the total distribution function $f_{\bm{k}}$ about the the equilibrium Fermi-Dirac distribution $f_{0}(\mathcal{E}_{\bm{k}})$ as
\begin{equation}
    f_{\bm{k}}
    =f_{0}(\mathcal{E}_{\bm{k}})+\delta\! f_{\bm{k}} 
\end{equation}
and keeping terms to linear order in $\delta\!f_{\bm{k}}$. 

The linearized Boltzmann equation can be recast as a system of linear equations \cite{Lucas2018,Cook2019,Qi2021}, a setup which we will use here extensively.
To do so, we introduce the following notation. 
We re-write the non-equilibrium distribution function as 
\begin{equation}\label{eq:psi}
    \delta\!f_{\bm{k}}
    =\fdd \psi_{\bm{k}}
\end{equation}
where $\psi_{\bm{k}}$ represents a non-equilibrium energy shift. Because of the singular behavior of $-\partial f_{0}/\partial\mathcal{E}_{\bm{k}}$, $\psi_{\bm{k}}$ is a smoother function of $\bm{k}$ than $\delta\!f_{\bm{k}}$, and it is standard to re-write the Boltzmann equation in terms of $\psi_{\bm{k}}$ \cite{Ziman1960}.
We represent the Bloch states using the ket $|\bm{k}{\rangle}$, and, reflecting the choice in \cref{eq:psi}, define the inner product 
\begin{equation}\label{eq:inner_product}
    {\langle}\bm{k}|\bm{k}'{\rangle}
    \equiv\fdd\delta_{\bm{k},\bm{k}'} .
\end{equation}
For a quantity $a_{\bm{k}}$, we define the vector
\begin{equation}\label{eq:vector}
    |a{\rangle}
    \equiv\sum_{\bm{k}}a_{\bm{k}}|\bm{k}{\rangle}
\end{equation}
and for a quantity $A_{\bm{k}\bm{k}'}$, we define the operator
\begin{equation}\label{eq:matrix}
    A|\bm{k}'{\rangle}
    \equiv\sum_{\bm{k}}A_{\bm{k}\bm{k}'}|\bm{k}{\rangle} .
\end{equation}
Note that as a result of our inner product definition, ${\langle}\bm{k}|a{\rangle}=(-\partial f_{0}/\partial\mathcal{E}_{\bm{k}})a_{\bm{k}}$ and ${\langle}\bm{k}|A|\bm{k}'{\rangle}=(-\partial f_{0}/\partial\mathcal{E}_{\bm{k}})A_{\bm{k}\bm{k}'}$.

Using these definitions, we can now write the linearized Boltzmann equation compactly as a system of linear equations:
\begin{equation}\label{eq:lin_boltz}
    (D+C)|\psi{\rangle}
    =-e\sum_{i}E_{i}|v_{i}{\rangle}
\end{equation}
where we have introduced the operator $D$ with
\begin{equation}
    D_{\bm{k}\bm{k}'}
    \equiv
    [\partial_{t}+\bm{v}_{\bm{k}}\cdot\nabla_{\bm{r}}]\,\delta_{\bm{k}\bm{k}'}
\end{equation}
and the linearized collision operator $C$ with
\begin{equation}
    C_{\bm{k}\bm{k}'}
    \equiv\left(\frac{\delta\mathcal{C}_{\bm{k}}}{\delta\!f_{\bm{k}'}}\right)_{\text{eq}}
\end{equation}
which arises from linearizing the collision operator about equilibrium and using that $\mathcal{C}_{\bm{k}}[f_{0}]=0$.

While our focus in this paper will mainly be on the use of phenomenological models for $C$, microscopically, it can be expressed as 
\begin{equation}\label{eq:Q}
    C_{\bm{k}\bm{k}'}
    =\frac{1}{f_{0}(\mathcal{E}_{\bm{k}'})(1-f_{0}(\mathcal{E}_{\bm{k}'}))}\left[
        -P_{\bm{k}\bm{k}'}
        +\sum_{\bm{k}''}P_{\bm{k}\bm{k}''}\delta_{\bm{k}\bm{k}'}
    \right] .
\end{equation}
Here $P_{\bm{k}\bm{k}'}$ is the equilibrium transition rate from $\bm{k}$ to $\bm{k}'$, which can be found for a given scattering mechanism using Fermi's golden rule. \footnote{The first term of \cref{eq:Q} can be understood as follows: In the linearized Boltzmann equation, the total rate of change to $\delta\!f_{\bm{k}}$ due to scattering is $\sum_{\bm{k}'}C_{\bm{k}\bm{k}'}\delta\!f_{\bm{k}'}$. So $C_{\bm{k}\bm{k}'}$ can be understood as the rate of scattering from $\bm{k}$ to $\bm{k}'$ if $\bm{k}'$ were empty---i.e. if $\delta\!f_{\bm{k}'}=-f_{0}(\mathcal{E}_{\bm{k}'})$, this induces a rate of change of $-C_{\bm{k}\bm{k}'}f_{0}(\mathcal{E}_{\bm{k}'})$ in $\delta\!f_{\bm{k}}$. This rate is related to the probability per unit time $P_{\bm{k}\bm{k}'}$ of this transition occurring in equilibrium, except that the latter also includes an extra factor $(1-f(\mathcal{E}_{\bm{k}'}))$ for the probability that $\bm{k}'$ is unoccupied. The second term of \cref{eq:Q} applies to diagonal elements, and represents the inverse lifetime of state $\bm{k}$ \cite{Allen1996}.}

The definitions in \cref{eq:inner_product,eq:vector,eq:matrix} also allow us to compactly represent products of the type
\begin{equation}
    {\langle}b|a{\rangle}
    =\sum_{\bm{k}}\fdd b_{\bm{k}}^{*}a_{\bm{k}}
\end{equation}
and
\begin{equation}
    {\langle}b|F|a{\rangle}
    =\sum_{\bm{k}\bm{k}'}\fdd b_{\bm{k}}^{*}F_{\bm{k}\bm{k}'}a_{\bm{k}'}
\end{equation}
which will often occur throughout this work. An important example is electrical current, given by 
\begin{equation}\label{eq:current}
    \bm{J}
    =-e{\langle}\bm{v}|\psi{\rangle} .
\end{equation}

\section{Phenomenological model for conserved quantities}

\subsection{Construction of collision operator}

The Boltzmann equation as written in \cref{eq:lin_boltz} describes a system of $N_{\bm{k}}$ equations where $N_{\bm{k}}$ is the number of eigenstates of the single-particle Hamiltonian.
Instead of using the basis of single-particle eigenstates, one can instead construct a collision operator directly in its eigenbasis. This provides a pathway for constructing simple, phenomenological collision operators. The approach is to single out a subset $R$ of $N_{R}$ eigenmodes for which the relaxation rates are set explicitly, while all other eigenmodes are assumed to relax at a shared rate $\gamma_{c}$. This approach has two advantages: (1) the solution of the Boltzmann equation in this case involves solving a set of linear equations of dimension $N_{R}$ rather than $N_{\bm{k}}$; (2) one can directly examine the consequences of the (quasi-)conservation of the eigenmodes in $R$ by setting $\gamma_{r,m}\ll\gamma_{c}$ for $m\in R$. A similar approach has been used for isotropic \cite{Guo2017} and anisotropic \cite{Cook2019,Qi2021,Valentinis2023} metals.

Let $\{|\chi_{m}{\rangle}\}$ be the complete set of eigenmodes of the collision operator with eigenvalues $\gamma_{m}$:
\begin{equation}
    C|\chi_{m}\rangle=\gamma_{m}|\chi_{m}\rangle
\end{equation}
The collision operator $C$ is Hermitian, and therefore its eigenvalues $\gamma_{m}$ are real.
We are interested in a simplified collision operator in which all modes are relaxed at a rate $\gamma_{c}$, except for a subset $R$ for which we will specify a distinct relaxation rate $\gamma_{r,m}$:
\begin{equation}
    \gamma_{m}
    =\begin{cases}
        \gamma_{r,m} & m\in R\\
        \gamma_{c} & \text{otherwise} 
    \end{cases}
\end{equation}
Using the completeness of the eigenbasis, the collision operator can then be written as 
\begin{equation}\label{eq:conserved_mode_collision_operator}
    C=\gamma_{c}-\sum_{m\in R}(\gamma_{c}-\gamma_{r,m})\frac{|\chi_{m}\rangle\!\langle\chi_{m}|}{\langle\chi_{m}|\chi_{m}\rangle} .
\end{equation}
Inserting our simplified collision operator into the Boltzmann equation (\cref{eq:lin_boltz}), we obtain
\begin{equation}\label{eq:Boltz_phenom}
    |\psi{\rangle}
    =\sum_{n{\in}R}
    \frac
        {\gamma_{c}-\gamma_{r,n}}
        {{\langle}\chi_{n}|\chi_{n}{\rangle}}
        M|\chi_{n}{\rangle}\!{\langle}\chi_{n}|\psi{\rangle}
    -e\sum_{i}E_{i}|v_{i}{\rangle}
\end{equation}
where we have defined $M\equiv(\gamma_{c}+D)^{-1}$.
Taking the product of ${\langle}\chi_{m}|$ with \cref{eq:Boltz_phenom} for each mode in $R$ yields a system of $N_{R}$ linear equations. Solving this system of equations for the products ${\langle}\chi_{m}|\psi{\rangle}$ and inserting the results into \cref{eq:Boltz_phenom} completes the solution of the Boltzmann equation.


\subsection{Solution for channel geometry}

We start by considering the general case of an electric field along $y$ which is spatially varying along $x$: $\bm{E}=E_{y}(x)\bm{\hat{y}}$. The general solution of \cref{eq:Boltz_phenom} is
\begin{equation}\label{eq:general}
    |\psi{\rangle}
    =|\psi_{c}{\rangle}
    +|\psi_{p}{\rangle}
\end{equation}
with the complementary solution 
\begin{equation}\label{eq:complementary}
    |\psi^{c}{\rangle}
    =|Ae^{-x\gamma_{c}/v_{x}}{\rangle}
\end{equation}
where the as-yet unspecified constant $A_{\bm{k}}$ is determined by the boundary conditions, and the particular solution is
\begin{equation}\label{eq:particular}
    |\psi^{p}{\rangle}
    =
    \int_{\{x_{0}\}}
        dx'|e^{-(x-x')\gamma_{c}/v_{x}}p(x'){\rangle}
\end{equation}
with 
\begin{equation}\label{eq:particular2}
    |p(x'){\rangle}
    \equiv
    \sum_{n\in X}
    \frac{\gamma_{c}-\gamma_{r,n}}{{\langle}\chi_{n}|\chi_{n}{\rangle}}
    |\chi_{n}/v_{x}{\rangle}\!{\langle}\chi_{n}|\psi{\rangle}
    -eE_{y}(x')|v_{y}/v_{x}{\rangle}
\end{equation}
and where $\{x_{0}\}$ depends on the sample's boundaries.

The case of a monochromatic electric field $E_{y}(x)=E_{0}e^{iqx}$, as considered in ref. \cite{Valentinis2023}, corresponds to 
\begin{equation}\label{eq:Mq}
    M(q,\omega)
    =\frac{1}{\gamma_{c}-i\omega+iv_{x}q}
\end{equation}
which follows from \cref{eq:general,eq:complementary,eq:particular,eq:particular2} by taking the range of integration $\{x_{0}\}$ to be $(-\infty,\infty)$. This could correspond to a theoretical scenario with complete translational invariance, or a semi-infinite sample with specular boundary scattering. The latter is because specular scattering in a sample occupying the domain $x>0$ can be described equivalently over the domain $(-\infty,\infty)$ by taking the current for $x<0$ as the reflection of that for $x>0$ \cite{Reuter1948}. Conveniently, the electrodynamics of a semi-infinite sample with diffuse boundary scattering can also be related to $\sigma(q,\omega)$ \cite{Dingle1953}.

Here we are interested in finding $M$ for the case of a channel carrying a DC current and with a finite width $W$. We take the current to be  along $y$ and the channel to extend from $x=-W/2$ to $x=W/2$. In this case, $D=v_{x}\partial_{x}$. 
For simplicity, we specialize to the case of a Fermi surface orientation relative to the channel which has mirror symmetry about $x=0$. As has been demonstrated experimentally in PdCoO$_2$ \cite{Bachmann2022}, if this mirror symmetry is broken this can give rise to a  transverse electric field along $x$. 
Under these assumptions, the electric field is spatially uniform: $\bm{E}=E_{0}\bm{\hat{y}}$.
The complementary solution is as in \cref{eq:complementary} while the particular solution in \cref{eq:particular} is simplified because the electric field is independent of $x$:
\begin{equation}
    |\psi^{(p)}{\rangle}
    =\sum_{n{\in}X}
    \frac{1-\gamma_{r,n}/\gamma_{c}}{{\langle}\chi_{n}|\chi_{n}{\rangle}}
    |\chi_{n}{\rangle}\!{\langle}\chi_{n}|\psi{\rangle}
    -\frac{eE_{0}}{\gamma_{c}}|v_{y}{\rangle}.
\end{equation}
Next we must apply boundary conditions to determine $A_{\bm{k}}$. We assume diffuse scattering of electrons from the boundaries:
%
%
%
\begin{equation}\label{eq:BC}
    |\psi^{\pm}(x=\mp W/2){\rangle}
    =0 
\end{equation}
where $|\psi^{+(-)}{\rangle}$ corresponds to the distribution function for electrons with $v_{\bm{k}x}>0$ ($v_{\bm{k}x}<0$).
Note that in the absence of mirror symmetry, the right-hand side of \cref{eq:BC} should be replaced by a constant determined by the condition that $J_{x}(x=\pm W/2)=0$. 
Applying \cref{eq:BC} gives
\begin{equation}
    |A{\rangle}=\left|\psi^{(p)}\exp\left(-\frac{W/2}{|v_{x}|/\gamma_{c}}\right)\right\rangle
\end{equation}
so that
\begin{equation}\label{eq:Mx}
    M(x)
    =\frac{1}{\gamma_{c}}
    \left[
        1-\exp\left(
            -\frac{x}{v_{x}/\gamma_{c}}
            -\frac{W/2}{|v_{x}|/\gamma_{c}}
        \right)
    \right] .
\end{equation}
Finally, we define a spatial average over the width of the channel as
\begin{equation}\label{eq:spatial_average}
    \overline{\mathcal{A}(x)}
    =\frac{1}{W}\int_{-W/2}^{W/2}dx\,\mathcal{A}(x)
\end{equation}
which we use to compute the channel-averaged conductivity $\overline{\sigma_{yy}(x)}$. We generalize \cref{eq:Mx} to include specular scattering in \cref{sec:BCs_general}, but note that for completely specular scattering the current in the channel is spatially uniform, and the conductivity is always equal to the bulk conductivity.

        
%

\subsection{Choice of conserved quantities}
Scattering in metals must conserve non-equilibrium particle number, so that $R$ must always contain the mode $|\chi_{n}{\rangle}=|1{\rangle}$ with associated eigenvalue $\gamma_{r,n}=0$. A common minimal model for comparing ohmic, hydrodynamic, and ballistic regimes is the Callaway dual-relaxation-time approximation (dRTA). In the Callaway dRTA, $R$ additionally includes each of the Cartesian components of ``momentum'', which are relaxed at a rate $\gamma_{r}$. While one can in principle include further modes in $R$, the Callaway dRTA will be our focus throughout much of the remainder of this paper.
While the Callaway dRTA was originally proposed in the context of the phonon Boltzmann equation \cite{Callaway1959}, it has recently been used extensively in the field of non-local electrical transport---as applied to isotropic, two-dimensional metals---both in theoretical work (e.g. refs. \cite{Guo2017,Scaffidi2017,Lucas2018}) and in the analysis of experimental data (e.g. refs. \cite{Molenkamp1994,DeJong1995,Moll2016,Vool2021}).
It is motivated by a situation in which there are two scattering sources, one with a rate $\gamma_{A}$ which only conserves particle number (often taken to be electron-impurity scattering) and with a rate $\gamma_{B}$ that conserves particle number and ``momentum'' (often taken to be normal electron-electron scattering). Then $\gamma_{r}=\gamma_{A}$ and $\gamma_{c}=\gamma_{A}+\gamma_{B}$. (The rate $\gamma_{A}$ contributes to both $\gamma_{r}$ and $\gamma_{c}$ because scattering mechanism A relaxes all eigenmodes of the collision operator other than particle number---see \cref{sec:convention_comparison} for a derivation and discussion of this relationship.)

Recently, Refs. \cite{Qi2021,Varnavides2022,Valentinis2023} have applied the Callaway dRTA to anisotropic metals.
While some authors have chosen crystal momentum as the conserved quantity \cite{Qi2021}, others have chosen momentum (or, equivalently, group velocity, since ${{\langle}\hat{\bm{p}}}{\rangle}_{\bm{k}}=m\bm{v}_{\bm{k}}$) \cite{Varnavides2022b,Valentinis2023}.
While both choices have merits, it is important to recognize the distinction between the two. 
A microscopic motivation for considering the crystal momentum-based scenario is that there are two scattering mechanisms for which crystal momentum is completely conserved: normal (i.e. non-Umklapp) electron-electron scattering and normal electron-phonon scattering under complete phonon drag. 
The resulting hydrodynamic equations in this case are conservation laws for number density and crystal momentum density. It is worth remembering that the viscosity entering these equations characterizes the transport of crystal momentum, whereas momentum is typically the quantity more directly accessible by experiment. The hydrodynamic equations in this scenario contain an additional ``incoherent conductivity'' term owing to the distinction between crystal momentum and group velocity \cite{Cook2019,Qi2021}, which would not be present in the corresponding equations for the momentum-based scenario. 
One way to motivate the momentum-based scenario is purely phenomenological: it provides a minimal model for scattering beyond that observable by local transport. Because $\gamma_{r,v}$ is exactly the scattering rate determining the local conductivity, $\gamma_{c}$ can be viewed as a single phenomenological parameter accounting for the additional scattering processes observable within non-local transport. 



\subsection{Solution to Boltzmann equation in dual relaxation-time approximation}

Here we apply the above model to the case that either crystal momentum or momentum is relaxed at a different rate than other non-equilibrium quantities.
In this case, we must solve the  following set of linear equations (which arise from taking the product of ${\langle}\chi_{m}|$ with \cref{eq:Boltz_phenom} for each of our three chosen eigenmodes):
\begin{widetext}
    \begin{equation}
        \begin{pmatrix}
            1-\frac{\gamma_{c}}{{\langle}1|1{\rangle}}{\langle}1|M|1{\rangle} 
            & -\frac{\gamma_{c}-\gamma_{r,\xi_{x}}}{{\langle}\xi_{x}|\xi_{x}{\rangle}}{\langle}1|M|\xi_{x}{\rangle} 
            & -\frac{\gamma_{c}-\gamma_{r,\xi_{y}}}{{\langle}\xi_{y}|\xi_{y}{\rangle}}{\langle}1|M|\xi_{y}{\rangle} \\
            -\frac{\gamma_{c}}{{\langle}1|1{\rangle}}{\langle}\xi_{x}|M|1{\rangle} 
            & 1-\frac{\gamma_{c}-\gamma_{r,\xi_{x}}}{{\langle}\xi_{x}|\xi_{x}{\rangle}}{\langle}\xi_{x}|M|\xi_{x}{\rangle} 
            & -\frac{\gamma_{c}-\gamma_{r,\xi_{y}}}{{\langle}\xi_{y}|\xi_{y}{\rangle}}{\langle}\xi_{x}|M|\xi_{y}{\rangle} \\
            -\frac{\gamma_{c}}{{\langle}1|1{\rangle}}{\langle}\xi_{y}|M|1{\rangle} 
            & -\frac{\gamma_{c}-\gamma_{r,\xi_{x}}}{{\langle}\xi_{x}|\xi_{x}{\rangle}}{\langle}\xi_{y}|M|\xi_{x}{\rangle} 
            & 1-\frac{\gamma_{c}-\gamma_{r,\xi_{y}}}{{\langle}\xi_{y}|\xi_{y}{\rangle}}{\langle}\xi_{y}|M|\xi_{y}{\rangle}
        \end{pmatrix}
        \begin{pmatrix}
            {\langle}1|\psi{\rangle} \\
            {\langle}\xi_{x}|\psi{\rangle} \\
            {\langle}\xi_{y}|\psi{\rangle}
        \end{pmatrix}
        =-eE_{y}\begin{pmatrix}
            {\langle}1|M|v_{y}{\rangle} \\
            {\langle}\xi_{x}|M|v_{y}{\rangle} \\
            {\langle}\xi_{y}|M|v_{y}{\rangle}
        \end{pmatrix}
    \end{equation}
where we have used a general variable $\xi$ which may be taken to be either $k$ for the crystal momentum case or $v$ for the momentum case.
If we assume two mirror planes, the system simplifies to
\begin{equation}
    \begin{pmatrix}
        1-\frac{\gamma_{c}}{{\langle}1|1{\rangle}}{\langle}1|M|1{\rangle}
        & -\frac{\gamma_{c}-\gamma_{r,\xi_{x}}}{{\langle}\xi_{x}|\xi_{x}{\rangle}}{\langle}1|M|\xi_{x}{\rangle} 
        & 0 \\
        -\frac{\gamma_{c}}{{\langle}1|1{\rangle}}{\langle}\xi_{x}|M|1{\rangle} 
        & 1-\frac{\gamma_{c}-\gamma_{r,\xi_{x}}}{{\langle}\xi_{x}|\xi_{x}{\rangle}}{\langle}\xi_{x}|M|\xi_{x}{\rangle}
        & 0 \\
        0
        & 0 
        & 1-\frac{\gamma_{c}-\gamma_{r,\xi_{y}}}{{\langle}\xi_{y}|\xi_{y}{\rangle}}{\langle}\xi_{y}|M|\xi_{y}{\rangle}
    \end{pmatrix}
    \begin{pmatrix}
        {\langle}1|\psi{\rangle} \\
        {\langle}\xi_{x}|\psi{\rangle} \\
        {\langle}\xi_{y}|\psi{\rangle}
    \end{pmatrix}
    =-eE_{y}\begin{pmatrix}
        0 \\
        0 \\
        {\langle}\xi_{y}|M|v_{y}{\rangle}
    \end{pmatrix}
\end{equation}
which gives ${\langle}1|\psi{\rangle}={\langle}\xi_{x}|\psi{\rangle}=0$ and 
\begin{equation}
    {\langle}\xi_{y}|\psi{\rangle}
    =\frac
        {-eE_{y}{\langle}\xi_{y}|M|v_{y}{\rangle}}
        {1-\frac{\gamma_{c}-\gamma_{r,\xi_{y}}}{{\langle}\xi_{y}|\xi_{y}{\rangle}}{\langle}\xi_{y}|M|\xi_{y}{\rangle}} .
\end{equation}
Then, using the definition of current in \cref{eq:current}, the conductivity for the crystal momentum case is 
\begin{equation}\label{eq:solution_k}
    \sigma_{yy}
    ={\langle}v_{y}|M|v_{y}{\rangle}
    +\frac{\gamma_{c}-\gamma_{r,k_{y}}}{{\langle}k_{y}|k_{y}{\rangle}}
    {\langle}v_{y}|M|k_{y}{\rangle}^{2}\left[
        1-\frac{\gamma_{c}-\gamma_{r,k_{y}}}{{\langle}k_{y}|k_{y}{\rangle}}{\langle}k_{y}|M|k_{y}{\rangle}
    \right]^{-1}
\end{equation}
\end{widetext}
and for the momentum case is
\begin{equation}\label{eq:solution_v}
    \sigma_{yy}
    ={\langle}v_{y}|M|v_{y}{\rangle}
    \left[
        1-\frac{\gamma_{c}-\gamma_{r,v_{y}}}{{\langle}v_{y}|v_{y}{\rangle}}{\langle}v_{y}|M|v_{y}{\rangle}
    \right]^{-1}
\end{equation}
where in \cref{eq:solution_k,eq:solution_v}, $M$ is given by \cref{eq:Mq} for a monochromatic electric field or by \cref{eq:Mx} for the channel geometry. 

In general, the spatial average \cref{eq:spatial_average} of the conductivity in the channel geometry must be taken over the entire expression in \cref{eq:solution_k} or \cref{eq:solution_v}. However, the case of a single relaxation-time approximation (sRTA)---i.e. $\gamma_{r}=\gamma_{c}$---allows for a particularly simple result:
\begin{equation}
    \overline{\sigma_{yy}^{\text{sRTA}}(x)}
    =e^{2}\overline{{\langle}v_{y}|M(x)|v_{y}{\rangle}}
    =e^{2}{\langle}v_{y}|\overline{M(x)}|v_{y}{\rangle}
\end{equation}
with
\begin{equation}
    \overline{M(x)}
    =\frac{1}{\gamma_{c}}\left\{1-\frac{|v_{x}|/\gamma_{c}}{W}\left[
        1
        -\exp\left(-\frac{W}{|v_{x}|/\gamma_{c}}\right)
    \right]\right\} .
\end{equation}

We take the opportunity here to review our results so far and their relation to the existing literature. We have derived closed-form solutions to the Callaway dRTA, valid for arbitrary Fermi surface geometry. \Cref{eq:solution_k,eq:solution_v} give the solutions for quasi-conserved crystal momentum and momentum, respectively, and apply to either a monochromatic electric field or the channel geometry depending on whether \cref{eq:Mq} or \cref{eq:Mx} is substituted for $M$.

Existing methods to solve the Callaway dRTA for the channel geometry, either for anisotropic \cite{Varnavides2022,Varnavides2022b} or isotropic metals \cite{Molenkamp1994,DeJong1995,Moll2016,Vool2021}, involved numerically solving the Boltzmann differential equation itself---here we have closed-form expressions which only involve numerically evaluating at most three integrals over the Fermi surface. Refs. \cite{Valentinis2023} and \cite{Qi2021}, whose derivations we followed closely, solved the Callaway dRTA for anisotropic metals only for the case of a monochromatic electric field, either for quasi-conserved momentum \cite{Valentinis2023} or crystal momentum \cite{Qi2021}.

\section{Comparison of circular and diamond Fermi surfaces}\label{sec:conductivity_results}

Here we wish to apply our model to examine the behavior of a nearest-neighbor tight-binding model on a square lattice at half filling, in the ``diamond'' and ``square'' Fermi surface orientations shown in \cref{fig:channels}. Before doing so, we briefly summarize the known results for the hydrodynamic prediction for an isotropic, viscous fluid, and the kinetic prediction from the Callaway dRTA for a circular Fermi surface. 

We define the ohmic, viscous, and ballistic regimes in \cref{tab:regimes} by the hierarchy of scales. To facilitate a comparison of length scales, we define the mean free paths $\lambda_{i}$ for $i\in\{r,c\}$ in terms of the corresponding scattering rates as $\lambda_{i}\equiv v_{F}/\gamma_{i}$, where $v_{F}$ is a thermally-averaged velocity magnitude: $v_{F}\equiv\sum_{\bm{k}}(-\partial f_{0}/\partial\mathcal{E}_{\bm{k}})|\bm{v}_{\bm{k}}|/\sum_{\bm{k}}(-\partial f_{0}/\partial\mathcal{E}_{\bm{k}})$.

Throughout this section, we will make a distinction between the \textit{regime} as defined purely by the hierarchy of scales and the actual \textit{behavior} of the conductivity. While these two classifications of transport are aligned for an isotropic Fermi surface, we shall see that the same is not always true for an anisotropic Fermi surface. 

While here we limit our discussion to defining the regimes by frequency- or length-scale, in an experimental setting the scattering rates $\gamma_{r}$ and $\gamma_{c}$ are tuned by temperature. Our model is agnostic regarding the microscopic scattering mechanism, treating $\gamma_{r}$ and $\gamma_{c}$ as phenomenological parameters. However, in \cref{sec:temperature} we discuss the role of temperature in tuning between transport regimes if we associate $\gamma_{r}$ and $\gamma_{c}$ with various relevant scattering mechanisms.

Throughout the remainder of this section, we will assume the degenerate limit $T{\ll}T_{F}$ such that $(-\partial f_{0}/\partial\mathcal{E}_{\bm{k}})\to\delta(\mathcal{E}_{\bm{k}}-\mathcal{E}_{F})$ and all sums of the type $\sum_{\bm{k}}(-\partial f_{0}/\partial\mathcal{E}_{\bm{k}})$ are restricted to the Fermi surface. For simplicity, when considering the wavevector-dependent conductivity, we will take $\omega=0$.

\begin{table}[!htbp]
    \begin{tabular}{p{2cm}p{3cm}p{3cm}}
        \toprule
        Regime & By frequencies & By lengths \\
        \midrule
        Ohmic 
        & $\sqrt{\gamma_{r}\gamma_{c}}\gg v_{F}q$
        & $\sqrt{\lambda_{r}\lambda_{c}}\ll W$ \\
        Viscous
        & $\sqrt{\gamma_{r}\gamma_{c}}\ll v_{F}q\ll \gamma_{c}$ 
        & $\lambda_{c}\ll W\ll\sqrt{\lambda_{r}\lambda_{c}}$ \\
        Ballistic
        & $v_{F}q\gg\gamma_{c}$
        & $W\ll\lambda_{c}$ \\
        \bottomrule
    \end{tabular}
    \caption{Definition of transport regimes by hierarchy of scales.}
    \label{tab:regimes}
\end{table}

\subsection{Viscous fluid}

As considered by Gurzhi \cite{Gurzhi1963,Gurzhi1968}, the hydrodynamic equation of motion for the velocity field $u$ of an isotropic, viscous, charged fluid is 
\begin{equation}
    (\nu\partial_{x}^{2}-\gamma_{r}+i\omega)u=-\frac{e}{m}E_{y}
\end{equation}
Then the channel-averaged conductivity for no-slip boundary conditions ($u(\pm W/2)=0$) is \footnote{
    See Ref. \cite{Levchenko2020} for a generalization of \cref{eq:hydro_ch} to arbitrary slip length, and Ref. \cite{Jaggi1991} for results for channels with different cross-sectional geometries.
} 
\begin{equation}\label{eq:hydro_ch}
    \begin{split}
        \overline{\sigma(x)}
        &=\dfrac{D}{\gamma_{r}}\left[1-\frac{l_{G}}{W/2}\tanh\frac{W/2}{l_{G}}\right]\\
        &=\begin{cases}
            \dfrac{D}{\gamma_{r}} & l_{G}\ll W \vspace{0.5em} \\
            \dfrac{D(W/2)^{2}}{3\nu} & l_{G}\gg W
        \end{cases}
    \end{split}
\end{equation}
and the wavevector-dependent conductivity is \cite{Forcella2014}
\begin{equation}\label{eq:hydro_q}
    \begin{split}
        \sigma(q)
        &=\dfrac{D}{\gamma_{r}+\nu q^{2}} \\
        &=\begin{cases}
            \dfrac{D}{\gamma_{r}} & l_{G}q\ll 1 \vspace{0.5em} \\
            \dfrac{D}{\nu q^{2}} & l_{G}q\gg 1
        \end{cases}
    \end{split}
\end{equation}
%
where $l_{G}\equiv\sqrt{\nu/\gamma_{r}}$ and $D\equiv ne^{2}/m$ where $n$ is the electron number density.

\subsection{Circular Fermi surface}

\begin{figure}[!htbp]
    \includegraphics{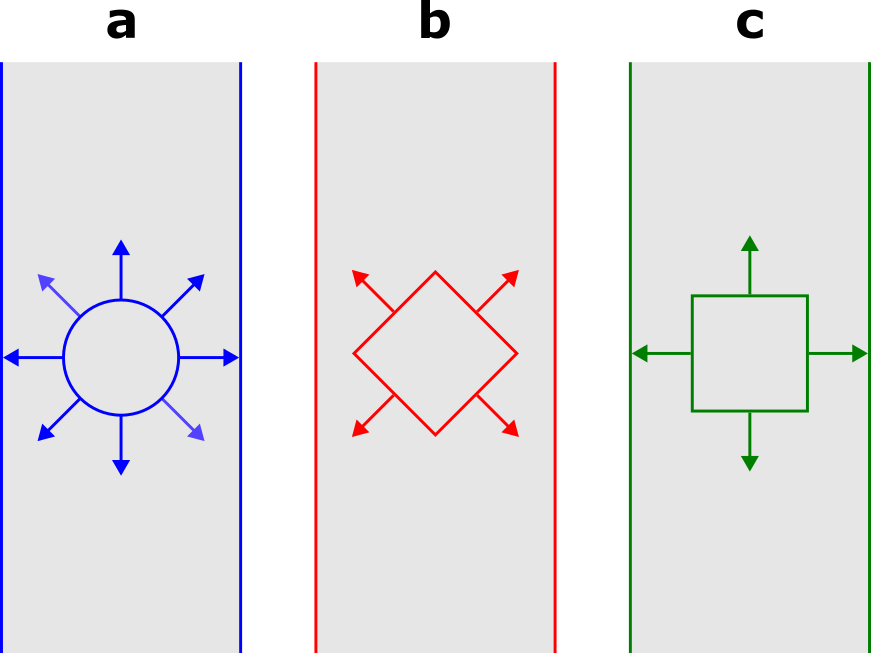}
    \caption{Channel orientations: (a) circular Fermi surface, (b) ``diamond'' orientation, and (c) ``square'' orientation.}
    \label{fig:channels}
\end{figure}

For a circular Fermi surface, the Callaway dRTA model gives \footnote{
    The ballistic result in \cref{eq:ci_ch} was found in Ref. \cite{Beenakker1988} within a single-rate relaxation-time approximation (sRTA), i.e. $\gamma_{c}=\gamma_{r}$. Results for the ballistic regime for a spherical Fermi surface (found within the sRTA) for this and other cross-sectional geometries are reviewed in Ref. \cite{Sondheimer2001}.
}
\begin{equation}\label{eq:ci_ch}
    \overline{\sigma(x)}
    =\begin{cases}
        \dfrac{D}{\gamma_{r}}
        & \text{ohmic regime} \vspace{0.5em} \\ 
        \dfrac{D\gamma_{c}W^{2}}{3v_{F}^{2}} \vspace{0.5em} 
        & \text{viscous regime}\\
        \dfrac{2D}{\pi v_{F}}W\ln(\lambda_{c}/W)
        & \text{ballistic regime}
    \end{cases}
\end{equation}
and
\begin{equation}\label{eq:ci_q}
    \sigma(q)
    =\begin{cases}
        \dfrac{D}{\gamma_{r}}
        & \text{ohmic regime} \vspace{0.5em} \\ 
        \dfrac{D}{(v_{F}q)^{2}/(4\gamma_{c})}
        & \text{viscous regime} \vspace{0.5em} \\
        \dfrac{D}{v_{F}q/2} 
        & \text{ballistic regime} .
    \end{cases}
\end{equation}
For convenience, we have introduced the Drude weight $D_{ii}\equiv{\langle}v_{i}|v_{i}{\rangle}$. For the three Fermi surface geometries considered here, $D_{xx}=D_{yy}\equiv D$.

By comparing \cref{eq:hydro_ch,eq:hydro_q} with \cref{eq:ci_ch,eq:ci_q}, we see that the Callaway dRTA result in the viscous regime matches the hydrodynamic result with the identification that the viscosity is given by \cite{Alekseev2016,Fritz2023} \footnote{
    The frequency-dependent shear viscosity arising from electron-electron interactions in a Galilean-invariant system in 2D or 3D is \cite{Abrikosov1959,Conti1999,Tokatly2000,Forcella2014,Levchenko2020,Valentinis2021-2}
    \begin{equation*}
        \nu(\omega)
        =\frac{v_{F}^{2}}{(2+d)(\gamma_{c}-i\omega)}
        \frac{1}{1+F_{1}^{S}/d} 
    \end{equation*}
    where $F_{1}^{S}$ is the first Landau parameter.
    However, in our model, including \cref{eq:ci_viscosity}, $\gamma_{c}$ is a phenomenological parameter which includes contributions from any scattering source that relaxes the eigenmodes of the collision operator that are orthogonal to momentum. This includes even electron-impurity scattering, as has been discussed by Ref. \cite{Alekseev2016}. See also the discussion in \cref{sec:callaway_rates,sec:convention_comparison}.
}
\begin{equation}\label{eq:ci_viscosity}
    \nu
    =\frac{v_{F}^{2}}{4\gamma_{c}} .
\end{equation}

\subsection{Diamond Fermi surface}

Here we consider a nearest-neighbor tight-binding model on a square lattice at half filling, which gives rise to a diamond Fermi surface. Analysis of the non-local transport associated with this simple Fermi surface serves as an excellent illustration of the subtleties introduced by the non-equivalence of crystal momentum and momentum.

We use the tight-binding dispersion in \cref{eq:tight_binding} not only to derive the Fermi surface geometry, but also to obtain the variation in magnitude and direction of the group velocity along the Fermi surface. In general the conductivities must be evaluated numerically, with analytic results only available in certain limiting cases.

\subsubsection{``Diamond'' orientation}

\begin{figure*}[!htbp]
    \includegraphics{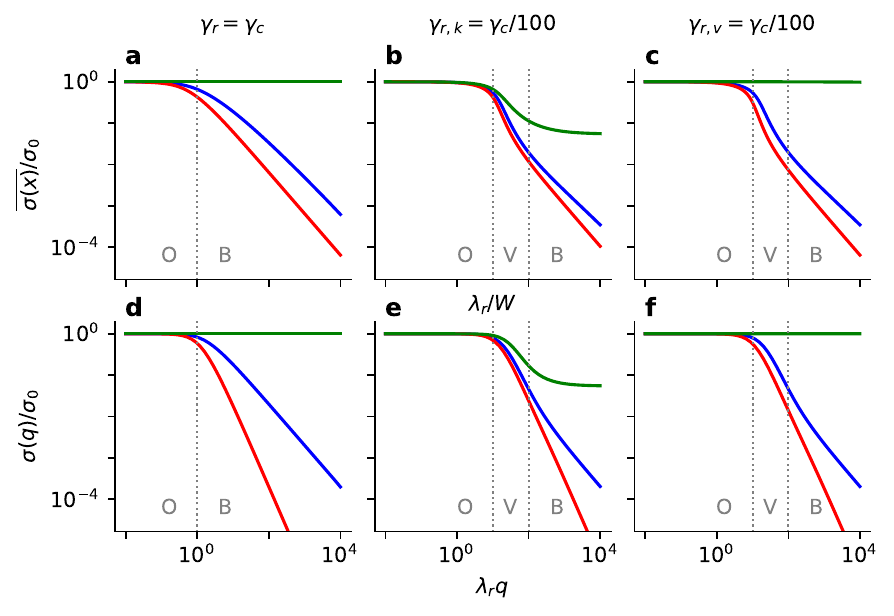}
    \caption{Effect of Fermi surface geometry, choice of quasi-conserved quantity, and experimental quantity. The top row shows average channel conductivity and the bottom shows wavevector-dependent conductivity, both normalized by the bulk (ohmic) conductivity $\sigma_{0}$. The left column corresponds to a single relaxation rate, the middle column to slow relaxation of total crystal momentum, and the right column to slow relaxation of total group velocity. The circular Fermi surface is in blue, ``diamond'' Fermi surface in red, and ``square'' Fermi surface in green, corresponding to the coloring in \cref{fig:channels}. The vertical dashed lines represent the crossovers between the \underline{o}hmic, \underline{v}iscous, and \underline{b}allistic regimes, as defined in \cref{tab:regimes}. First, consider the ``diamond'' Fermi surface. (d) Shows that the ``diamond'' Fermi surface exhibits $\sigma(q)\sim q^{-2}$ in region B as opposed to the $\sigma(q)\sim q^{-1}$ for a circular Fermi surface. This power law is typically associated with viscous behavior (see region V in (b), (c), (e) and (f)), even though in panel (d) no (crystal) momentum-conserving scattering has been introduced. Meanwhile, (a) shows that the channel-averaged conductivity of the ``diamond'' Fermi surface only deviates moderately from that of the circular Fermi surface. In all plots, the behavior of the ``diamond'' Fermi surface in region B shows a breakdown in the correspondence between $\sigma(q)$ and $\overline{\sigma(x)}$. Next, consider the ``square'' Fermi surface. In all panels, we see that the conductivity of the ``square'' is constant in region B. This behavior is typically associated with the ohmic regime, despite the hierarchy of length scales implying a ballistic regime. In (b), (c), (e), and (f), we see that in both $\overline{\sigma(x)}$ and $\sigma(q)$, whether crystal momentum or momentum is being slowly relaxed leads to different behavior for the ``square'' Fermi surface. For slow crystal momentum relaxation ((b) and (e)), the conductivity shows scale dependence in region V, while for slow momentum relaxation ((c) and (f)), this behavior is suppressed. In all cases, the conductivity has a constant asymptote in region B, while in (b) and (e) this constant value is lower than the bulk conductivity.}
    \label{fig:results}
\end{figure*}

Here we consider the case in which the channel is aligned with the crystallographic axes, as shown in \cref{fig:channels}b. We find that in the ballistic regime
\begin{equation}
    \overline{\sigma(x)}
    \approx\frac{DW}{\sqrt{2}v_{F}}
\end{equation}
and
\begin{equation}\label{eq:di_ballistic}
    \sigma(q)
    \approx\frac{D}{(v_{F}q)^{2}/(2\gamma_{c})} .
\end{equation}
These results hold independently of whether crystal momentum or group velocity is slowly relaxed, and in fact independently of the existence of a slowly-relaxed quantity---i.e. they hold for a single-relaxation-time approximation where $\gamma_{c}=\gamma_{r}\equiv\gamma$. 
These results are surprising in two ways.

The first surprise is that the behavior of $\sigma(q)$ in the ballistic regime is qualitatively different from that for a circular Fermi surface. Here, in a region defined as ballistic based on the relative magnitudes of the scales involved, the apparent behavior is viscous (\cref{fig:results}d)---even in the absence of momentum-conserving scattering. Comparing \cref{eq:di_ballistic} with the isotropic hydrodynamic result (\cref{eq:hydro_q}) yields an apparent viscosity of 
\begin{equation}
    \nu
    =\frac{v_{F}^{2}}{2\gamma_{c}}
\end{equation}
This apparent viscosity in the ballistic regime for the diamond Fermi surface is twice as large as the viscosity in the viscous regime for the circular Fermi surface (cf. \cref{eq:ci_viscosity}).
The effect of this behavior would be experimentally observable via surface impedance measurements \cite{Valentinis2023}. (\Cref{sec:impedance} discusses the relevant parameter range for detecting the behavior of $\sigma(q)$ via the frequency-dependent surface impedance.)

The second surprise is that the behaviors of $\sigma(q)$ and $\overline{\sigma(x)}$ do not match one another. Unlike $\sigma(q)$, the behavior of $\overline{\sigma(x)}$ is only slightly modified relative to that of a circular Fermi surface---$\sim W$ versus $\sim W\log(\lambda_{c}/W)$ (\cref{fig:results}a). Therefore, $\sigma(q)$ would appear ``viscous'' while $\overline{\sigma(x)}$ would appear ``ballistic''. 

\subsubsection{``Square'' orientation}

Here we consider the case in which the channel is rotated by 45$^{\circ}$ relative to the crystallographic axes, as shown in \cref{fig:channels}c. We find that in the ballistic regime, both $\overline{\sigma(x)}$ and $\sigma(q)$ are constant. In a single-relaxation-time approximation or for slow relaxation of total group velocity, 
\begin{equation}
    \overline{\sigma(x)}(W)
    =\sigma(q)
    =\frac{D}{\gamma_{r,v}}
\end{equation}
for all values of $W$ or $q$, respectively (\cref{fig:results}a, d, c, \& f).
Meanwhile, for slow relaxation of total crystal momentum, the constant values of $\overline{\sigma(x)}$ and $\sigma(q)$ in the ballistic regime are lower than those in the ohmic regime and depend on the value of $\gamma_{c}$ (\cref{fig:results}b \& e).
Once again, qualitatively new behavior relative to a circular Fermi surface emerges. The ballistic regime is suppressed because those electrons that contribute to the conductivity propagate down the length of the channel without colliding with the boundaries, even when the channel is narrower than the magnitude of the mean free path. The fact that scale-dependent behavior in the viscous regime remains while ballistic behavior is suppressed (\cref{fig:results}b \& e) is an interesting demonstration of the different physics of these regimes as well as the difference between crystal momentum and group velocity in anisotropic metals. It is perhaps counterintuitive at first that the viscous regime should exist in this geometry---how can the flow in adjacent layers be coupled if the group velocity indicates that electrons in adjacent layers propagate parallel to one another? This is a manifestation of the fact that we have enforced a slower rate of \textit{crystal momentum} relaxation, and the crystal momentum does vary between adjacent layers. If we instead enforce a slower rate of \textit{momentum} relaxation, the viscous regime is suppressed as well (\cref{fig:results}c \& f).

\section{Single-particle, transport, and viscous scattering rates}

Here we wish to address the question of what scattering processes determine the local conductivity and viscosity tensors. In this section we are interested in bulk properties, independent of a specific geometry. For isotropic metals, the answer is known. It has been used to great advantage \cite{Guo2017,Ledwith2019b,Fritz2023} that the eigenfunctions of the collision operator for an isotropic, two-dimensional metal are the angular harmonics $\chi_{m}\sim e^{im\phi}$. In this case the scattering rate entering the local resistivity is the eigen-rate $\gamma_{1}$ for the $m=\pm1$ harmonics and that entering the local viscosity is the eigen-rate $\gamma_{2}$ for the $m=\pm2$ harmonics. In other words, the local resistivity and viscosity are related to the relaxation of $m=\pm1$ and $m=\pm2$ deformations of the Fermi surface. However, in anisotropic metals, these quantities do not necessarily correspond to eigenfunctions of the collision operator. Nor are the eigenfunctions of the collision operator angular harmonics, or in fact known at all. In \cref{sec:rate_defns}, we seek to define precisely the scattering rates that determine the local conductivity and viscosity tensors in anisotropic metals, and their relations to microscopic transition probabilities. We use a variational principle to find approximations to these scattering rates that are valid for arbitrary collision operator. The variational expression for the conductivity is known \cite{Allen1996}; we obtain that for the viscosity via generalization. In \cref{sec:callaway_rates}, we examine these scattering rates in the specific context of the Callaway dRTA model for the collision operator and discuss the implications. In particular, we show that while crystal momentum-conserving does not influence the local conductivity in isotropic metals, the same is not true of anisotropic metals. We present a simple model for the contribution of normal electron-electron scattering to the transport scattering rate for anisotropic metals.

\subsection{General definitions}\label{sec:rate_defns}

Following ref. \cite{Allen1996}, we define a generalized weighted scattering rate as 
\begin{equation}\label{eq:weighted_scattering}
    \Gamma[w_{\bm{k}\bm{k}'}]
    =N(0)\frac
        {\frac{1}{k_{B}T}\sum_{\bm{k}\bm{k}'}P_{\bm{k}\bm{k}'}w_{\bm{k}\bm{k}'}}
        {\sum_{\bm{k}\bm{k}'}w_{\bm{k}\bm{k}'}(-\partial f_{0}/\partial\mathcal{E}_{\bm{k}})(-\partial f_{0}/\partial\mathcal{E}_{\bm{k}'})} .
\end{equation}
The scattering rates corresponding to various physical quantities can be expressed using \cref{eq:weighted_scattering} given a corresponding weighting function $w_{\bm{k}\bm{k}'}$.
Physically, this describes how different quantities are more or less sensitive to different scattering events.

\subsubsection{Single-particle scattering rate}

The single-particle scattering rate for state $\bm{k}$ is given by
\begin{equation}
    \gamma_{\text{sp}}(\bm{k})
    =\frac{1}{f_{\bm{k}}^{0}(1-f_{\bm{k}}^{0})}\sum_{\bm{k}'}P_{\bm{k}\bm{k}'}
\end{equation}
where $f_{\bm{k}}^{0}$ is the Fermi-Dirac function and $P_{\bm{k}\bm{k}'}$ is the equilibrium transition rate. In order to compare with other scattering rates, it is useful to define a thermally-averaged single-particle scattering rate:
\begin{equation}
    \gamma_{\text{sp}}
    =\frac{1}{N(0)}\sum_{\bm{k}}\gamma_{\text{sp}}(\bm{k})\left(-\frac{\partial f_{0}}{\partial\mathcal{E}_{\bm{k}}}\right) .
\end{equation}
This rate corresponds to the generalized rate with weighting function $1$:
\begin{equation}
    \gamma_{\text{sp}}
    =\Gamma[w_{\bm{k}\bm{k}'}=1] .
\end{equation}

\subsubsection{Transport scattering rate}

The bulk, DC conductivity is given by 
\begin{equation}
    \sigma_{ij}
    ={\langle}v_{i}|C^{-1}|v_{j}{\rangle}. 
\end{equation}
This may be written in the Drude form 
\begin{equation}
    \sigma_{ij}
    =\frac{D_{\sigma,ij}}{\gamma_{\sigma,ij}}
\end{equation}
if we define the Drude weight by 
\begin{equation}
    D_{\sigma,ij}\equiv{\langle}v_{i}|v_{j}{\rangle}
\end{equation}
and the transport scattering rate by 
\begin{equation}
    \frac{1}{\gamma_{\sigma,ij}}
    \equiv\frac{{\langle}v_{i}|C^{-1}|v_{j}{\rangle}}{{\langle}v_{i}|v_{j}{\rangle}} .
\end{equation}
In terms of the eigenvalues $\gamma_{m}$ of the collision operator, it is given by
\begin{equation}
    \frac{1}{\gamma_{\sigma,ij}}
    =\sum_{m}
    \frac{1}{\gamma_{m}}
    \frac
        {{\langle}v_{i}|\chi_{m}{\rangle}\!{\langle}\chi_{m}|v_{j}{\rangle}}
        {{\langle}v_{i}|v_{j}{\rangle}{\langle}\chi_{m}|\chi_{m}{\rangle}} .
\end{equation}
For a circular Fermi surface, the only non-zero overlaps are with the angular harmonic eigenfunctions with $m=\pm1$ and the viscous scattering rate is exactly the eigenvalue $\gamma_{1}$.
The lowest-order variational approximation for $\gamma_{\sigma,ij}$ is given by 
\begin{equation}
    \gamma_{\sigma,ij}^{(0)}
    =\Gamma[w_{\bm{k}\bm{k}'}=(v_{i}-v_{i}')(v_{j}-v_{j}')] .
\end{equation}
For an isotropic Fermi surface (in either 2 or 3 dimensions) and for the diagonal elements of the conductivity, this weighting factor reduces to $w_{\bm{k}\bm{k}'}\sim1-\cos\theta$ where $\theta$ is the scattering angle, and we recover the oft-cited weighting factor differentiating the single-particle and transport scattering rates \cite{Allen1996}. However, we emphasize that, contrary to the practice of using $w_{\bm{k}\bm{k}'}\sim1-\cos\theta$ for any Fermi surface \cite{Varnavides2022}, the correct weighting factor for an anisotropic Fermi surface is instead $(v_{i}-v_{i}')^{2}$.

\subsubsection{Viscous scattering rate}

The local crystal viscosity---named as such because it characterizes the flow of crystal momentum rather than momentum---can be written as 
\begin{equation}
    \eta_{ijkl}
    ={\langle}v_{i}k_{j}|C^{-1}|v_{k}k_{l}{\rangle}.
\end{equation}
In analogy with the conductivity, we write this as 
\begin{equation}
    \eta_{ijkl}
    =\frac{D_{\eta,ijkl}}{\gamma_{\eta,ijkl}}
\end{equation}
with
\begin{equation}
    D_{\eta,ijkl}
    \equiv{\langle}v_{i}k_{j}|v_{k}k_{l}{\rangle}
\end{equation}
and
\begin{equation}
    \gamma_{\eta,ijkl}
    \equiv\frac
        {{\langle}v_{i}k_{j}|C^{-1}|v_{k}k_{l}{\rangle}}
        {{\langle}v_{i}k_{j}|v_{k}k_{l}{\rangle}} .
\end{equation}
In terms of the eigenvalues $\gamma_{m}$ of the collision operator, the viscous scattering rate
\begin{equation}
    \frac{1}{\gamma_{\eta,ijkl}}
    =\sum_{m}
    \frac{1}{\gamma_{m}}
    \frac
        {{\langle}v_{i}k_{j}|\chi_{m}{\rangle}\!{\langle}\chi_{m}|v_{k}k_{l}{\rangle}}
        {{\langle}v_{i}k_{j}|v_{k}k_{l}{\rangle}\!{\langle}\chi_{m}|\chi_{m}{\rangle}} .
\end{equation}
For a circular Fermi surface, the only non-zero overlaps are with the angular harmonic eigenfunctions with $m=\pm2$ and the viscous scattering rate is exactly the eigenvalue $\gamma_{2}$. 
The lowest-order variational approximation for $\gamma_{\eta,ijkl}$ is given by 
\begin{equation}
    \gamma_{\eta,ijkl}^{(0)}
    =\Gamma[w_{\bm{k}\bm{k}'}=(v_{i}k_{j}-v_{i}'k_{j}')(v_{k}k_{l}-v_{k}'k_{l}')] .
\end{equation}
The above analysis may be repeated replacing $k_{i}$ by $v_{i}$ to evaluate the viscosity (rather than crystal viscosity) and its associated scattering rate.

\subsection{Callaway model}\label{sec:callaway_rates}

Within the Callaway dRTA model with rate $\gamma_{r,k}$ for crystal momentum and $\gamma_{c}$ otherwise, we have
\begin{equation}
    \gamma_{\text{sp}}
    =\gamma_{c}
\end{equation}
and
\begin{equation}\label{eq:transport_rate_dRTA}
    \frac{1}{\gamma_{\sigma}}
    =w\frac{1}{\gamma_{r,k}}+(1-w)\frac{1}{\gamma_{c}}
\end{equation}
where 
\begin{equation}\label{eq:transport_weight_dRTA}
    w\equiv\frac{{\langle}v_{y}|k_{y}{\rangle}^{2}}{{\langle}v_{y}|v_{y}{\rangle}{\langle}k_{y}|k_{y}{\rangle}}
\end{equation}
and
\begin{equation}
    \gamma_{\eta}
    =\gamma_{c}.
\end{equation}

It is worth explicitly reflecting on the meaning of these results. It is often stated that ``momentum''-conserving scattering does not contribute to resistivity. However, this statement is not true for \textit{crystal momentum}-conserving scattering (such as normal electron-electron scattering) and an anisotropic Fermi surface. While this (often underappreciated) fact has already been reported \cite{Pal2012}, the above results allow for a particularly transparent demonstration.
%
%

These simple expressions state that, at the level of the Callaway dRTA, the bulk conductivity can be written in the regular Drude form $\sigma_{yy}=D_{\sigma,yy}/\gamma_{\sigma,yy}$, except that the transport scattering rate $\gamma_{\sigma,yy}$ must be interpreted as a weighted average of the crystal momentum-relaxing and crystal momentum-conserving scattering rates (\cref{eq:transport_rate_dRTA}). The weighting function $w$ (\cref{eq:transport_weight_dRTA}) is a measure of the similarity of momentum and crystal momentum. For a circular Fermi surface $w=1$, and the transport (momentum-relaxing) scattering rate is exactly the crystal momentum-relaxing scattering rate. The quantity $w$ is exactly that plotted in \cref{fig:kv_overlap} for a tight-binding model on a square lattice as a function of Fermi energy.

Consider a simple model of electron-impurity scattering and electron-electron scattering. We take the electron-impurity scattering to be characterized by only a single rate $\gamma_{ei}$. We take electron-electron scattering to be characterized by two rates: normal electron-electron scattering which conserves crystal momentum, at a rate $\gamma_{ee}^{N}=(1-U)\gamma_{ee}$, and Umklapp electron-electron scattering which does not conserve crystal momentum, at a rate $\gamma_{ee}^{U}=U\gamma_{ee}$ where $U$ is the Umklapp efficiency. Then we may apply the Callaway dRTA with $\gamma_{r}=\gamma_{ei}+U\gamma_{ee}$ (assuming Matthiessen's rule) and $\gamma_{c}=(1-U)\gamma_{ee}$. Then the transport scattering rate is given by
\begin{equation}\label{eq:Callaway_transport_scattering}
    \frac{1}{\gamma_{\sigma}}
    =\frac{1}{\gamma_{ei}+U\gamma_{ee}}
    +(1-w)\frac{1}{(1-U)\gamma_{ee}}
\end{equation}
This formula gives a simple estimate for the contribution of normal electron-electron scattering to the transport scattering rate. We have shown how $w$ can change as a function of Fermi surface geometry in the context of a square lattice; \cref{eq:Callaway_transport_scattering} shows that for any deviation of $w$ from unity, normal electron-electron scattering contributes to the transport scattering rate. 
A fuller model would also include how the Umklapp efficiency $U$ evolves with Fermi surface geometry and filling. Nonetheless, another point becomes obvious from \cref{eq:Callaway_transport_scattering}: if the Umklapp efficiency is sufficiently high, then no level of electron-electron scattering can give rise to a large imbalance between the momentum-relaxing and momentum-conserving rates. This means that it is simply incorrect to attribute viscous behavior seen in any large Fermi surface metal to electron-electron scattering.

\section{Conclusions \& outlook}

Here we have examined a generalization of the Callaway dual-relaxation-time approximation (dRTA) model to anisotropic metals. We have expanded on previous work \cite{Qi2021,Varnavides2022,Varnavides2022b,Valentinis2023} to solve the Callaway dRTA in closed form for both the wavevector-dependent conductivity as well as for the conductivity of a channel with diffuse boundary scattering, and for slow relaxation of either crystal momentum or momentum. Furthermore, we have called attention to various conceptual issues unique to anisotropic metals. 
We have shown examples of how Fermi surface anisotropy and boundary conditions can lead to qualitatively different behaviors that confound the diagnosis of the underlying transport regime. Therefore, it is vital these factors are included in any analysis being used to interpret experimental data.

More broadly, we suggest a careful examination of the current paradigm in the field of non-local transport in ultra-pure metals, in which the focus is to classify transport as ohmic, hydrodynamic, or ballistic. A central feature of non-local transport is the coupling of different modes. In this way, a non-local transport measurement contains more information about microscopic scattering processes than a local one. While it is common practice to make a single-relaxation-time approximation (sRTA) when analyzing local transport properties (with the acknowledgment that the scattering rates for different quantities will differ), an sRTA cannot describe a non-local transport measurement unless the lifetimes of all coupled modes happen to be identical (e.g., if the only scattering source is point-like defects). There is already considerable interest in condensed matter physics at the information that can be gleaned by comparing different scattering rates---e.g. those from electrical and thermal conductivities---and the single particle rate. In non-local transport, a single measurement is already sensitive to multiple lifetimes. While a Callaway dRTA can give rise to ohmic, hydrodynamic, and ballistic regimes, full collision operators may give rise to a more rich landscape in between the ohmic and ballistic limits. The appeal of studying hydrodynamics likely comes from (1) the advantage of a simple, universal description of transport and (2) analogies with other fields of physics. However, the use of the dRTA may have risks: theoretically, other physics may be overlooked; experimentally, a Callaway dRTA may describe data better than an sRTA because it better approximates the structure of the full collision operator, even when scattering does not give rise to any conservation law. Where calculations using a full collision operator are possible, it will be interesting to compare with the Callaway dRTA. Some results are already available: results from randomly-generated collision operators suggest that the Callaway dRTA often performs well \cite{Varnavides2022b}; calculations for electron-phonon scattering in isotropic metals reveal a hierarchy of lifetimes \cite{Levchenko2020,Huang2021}; it has been shown that two distinct rates arise from electron-electron scattering on polygonal Fermi surfaces, leading to a failure of the Callaway dRTA at the ballistic-to-hydrodynamic crossover \cite{Cook2019}; perhaps most strikingly, calculations for normal electron-electron scattering in 2DEGs---the scattering mechanism which originally inspired the use of the Callaway dRTA---show that the eigenfrequencies of the collision operator in fact display a rich structure \cite{Ledwith2019a} so that the Callaway dRTA fails to correctly predict transport properties in this context \cite{Ledwith2019b}. Further analysis of full collision operators for different scattering mechanisms and Fermi surface geometries will be interesting, and may lead to the prediction of novel and testable phenomena. Gurzhi's famous work should therefore be regarded as the foundation of a much larger field than ``simple'' electron hydrodynamics.

\section*{Acknowledgments}

We acknowledge useful discussions with Thomas Scaffidi and Andrew Lucas. Research in Dresden benefits from the environment provided by the DFG Cluster of Excellence ct.qmat EXC 2147, project ID 390858940. D.V. acknowledges partial support by the European Commission's Horizon 2020 RISE program Hydrotronics (Grant No. 873028).

\appendix

\section{Comparison of scattering rate conventions: by mechanism vs. by eigenmode}\label{sec:convention_comparison}

In the existing literature, two slightly different conventions for the definition of scattering rates in the dRTA have been used. One groups scattering by the eigenmodes of the collision operator, the other groups scattering by the mechanism. Here we clarify the relationship between the two conventions. The former convention was used in the present work and also in Ref. \cite{Valentinis2023}. To the best of our knowledge, the latter convention was first used in the context of the electronic Boltzmann equation by \citet{DeJong1995}, and so we will refer to it as the ``deJM'' convention. It has also been used, e.g., by Refs. \cite{Moll2016,Scaffidi2017}. 

%

Consider Eq. 9 of Ref. \cite{DeJong1995}, which describes the contribution of a momentum-relaxing (MR) scattering mechanism to the collision integral for a 2DEG:
\begin{equation}\label{eq:dJM_ei}
    \left.\frac{\partial\psi(\phi)}{\partial t}\right|_{\text{MR}}
    =-\gamma_{\text{MR}}^{\text{dJM}}\,\psi(\phi)
    +\gamma_{\text{MR}}^{\text{dJM}}\int_{0}^{2\pi}\frac{d\phi'}{2\pi} \psi(\phi')
\end{equation}
Using our bra-ket notation, this can be written for arbitrary electronic dispersion as
\begin{equation}\label{eq:collision_ei}
    C_{\text{MR}}|\psi{\rangle}
    =\gamma_{\text{MR}}^{\text{dJM}}
    \left(
            1
            -\frac
                {|1{\rangle}\!{\langle}1|}
                {{\langle}1|1{\rangle}}
    \right)|\psi{\rangle} .
\end{equation}
(Note that the factor of $1/2\pi$ appearing in \cref{eq:dJM_ei} arises from applying \cref{eq:collision_ei} to a 2DEG, but is not the most general case.)

Consider Eq. 11 from Ref. \cite{DeJong1995}, which describes the contribution of a momentum-conserving (MC) scattering mechanism to the collision integral for a 2DEG: 
\begin{equation}\label{eq:dJM_ee}
    \begin{split}
        \left.\frac{\partial\chi(\phi)}{\partial t}\right|_{\text{MC}}
        =&-\gamma_{\text{MC}}^{\text{dJM}}\,\chi(\phi)\\
        &+\gamma_{\text{MC}}^{\text{dJM}}\int_{0}^{2\pi}\frac{d\phi'}{2\pi}\chi(\phi')[1+2\,\bm{v}'\cdot\bm{v}] .
    \end{split}
\end{equation}
Using our bra-ket notation, this can be written for arbitrary electronic dispersion as 
\begin{equation}\label{eq:collision_ee}
    C_{\text{MC}}|\psi{\rangle}
    =\gamma_{\text{MC}}^{\text{dJM}}\left(
        1
        -\frac
            {|1{\rangle}\!{\langle}1|}
            {{\langle}1|1{\rangle}}
        -\sum_{i=x,y}\frac
            {|\xi_{i}{\rangle}\!{\langle}\xi_{i}|}
            {{\langle}\xi_{i}|\xi_{i}{\rangle}}
    \right)|\psi{\rangle} .
\end{equation}
Here we have generalized to a variable $\xi$ which can be taken to be either crystal momentum or group velocity---the distinction is moot in the 2DEG case considered in  Ref. \cite{DeJong1995}.
(Note that the factors of $1/2\pi$ and $2$ appearing in \cref{eq:dJM_ee} arise from applying \cref{eq:collision_ee} to a 2DEG, but are not the most general case.) 

With the total collision operator as $C=C_{\text{MR}}+C_{\text{MC}}$, we have that 
\begin{equation}\label{eq:dJM_rates}
    \begin{split}
        C
        =&\;\gamma_{\text{MR}}^{\text{dJM}}+\gamma_{\text{MC}}^{\text{dJM}}\\
        &-(\gamma_{\text{MR}}^{\text{dJM}}+\gamma_{\text{MC}}^{\text{dJM}})\frac{|1{\rangle}\!{\langle}1|}{{\langle}1|1{\rangle}}\\
        &-\sum_{i}\gamma_{\text{MC}}^{\text{dJM}}\frac{|v_{i}{\rangle}\!{\langle}v_{i}|}{{\langle}v_{i}|v_{i}{\rangle}}
    \end{split}
\end{equation}
Upon comparison of \cref{eq:dJM_rates,eq:conserved_mode_collision_operator}, we see that the two conventions are equivalent with the identification that 
\begin{equation}\label{eq:gammar_dJM}
    \gamma_{r,\xi}=\gamma_{\text{MR}}^{\text{dJM}}
\end{equation}
and 
\begin{equation}\label{eq:gammac_dJM}
    \gamma_{c}=\gamma_{\text{MR}}^{\text{dJM}}+\gamma_{\text{MC}}^{\text{dJM}} .
\end{equation}

An intuitive understanding of the correspondence comes from considering how the scattering is grouped in the two conventions. \Cref{eq:collision_ei} shows that the dJM MR mechanism relaxes all eigenmodes at the rate $\gamma_{\text{MR}}^{\text{dJM}}$ (except for particle number), and \cref{eq:collision_ee} shows that the dJM MC mechanism relaxes all eigenmodes at the rate $\gamma_{\text{MC}}^{\text{dJM}}$ except for momentum (and particle number). It then follows that momentum relaxes at a rate $\gamma_{\text{MR}}^{\text{dJM}}$ (\cref{eq:gammar_dJM}) and that all other modes relax at a rate $\gamma_{\text{MR}}^{\text{dJM}}+\gamma_{\text{MC}}^{\text{dJM}}$ (\cref{eq:gammac_dJM}) (excluding particle number, which does not relax).

\section{The role of temperature}\label{sec:temperature}

As seen in \cref{tab:regimes}, the viscous regime is an intermediate-scale phenomenon. Given that the mean free paths/scattering rates used to define the regimes in \cref{tab:regimes} are almost always monotonic functions of temperature, the viscous regime is expected to typically occur within an intermediate temperature window. However, the detailed definition of this temperature window is not universal, but rather depends on the magnitudes and temperature dependences of the scattering rates specific to a given material. One of the motivations for the phenomenological model of the collision operator employed in this work is that it is agnostic to a particular scattering mechanism. Nonetheless, we will comment briefly here on a few of the most relevant scenarios.

In Gurzhi's earliest work on electron hydrodynamics \cite{Gurzhi1963}, he considered the temperature-dependent resistivity of a channel of width $W$ for which $\gamma_{c}$ is dominated by electron-electron scattering ($\gamma_{c}(T)=A_{c,2}T^{2}$). In this case the ballistic--viscous crossover occurs at a temperature $T_{b\leftrightarrow v}=\sqrt{\frac{v_{F}}{WA_{c,2}}}$. For $\gamma_{r}(T)=A_{r,n}T^{n}$, the viscous--ohmic crossover occurs at a temperature $T_{v\leftrightarrow c}=\frac{1}{A_{r,n}}\left(\frac{v_{F}^{2}}{W^{2}A_{c,2}}\right)^{1/n}$. The two scenarios considered by Gurzhi were $n=0$, as occurs for electron-impurity scattering, and $n=5$, as conventionally occurs for electron-phonon scattering in three-dimensional metals over the relevant temperature range. The relevant exponent for electron-phonon scattering is instead $n=1$ in (Al,Ga)As-based 2DEGs \cite{Kawamura1992} and mono- and bi-layer graphene \cite{Ho2018}.

It is also possible that $\gamma_{c}$ is itself dominated by electron-phonon scattering if the total momentum of the combined electron-phonon system is conserved. In this case, one would expect $\gamma_{c}(T)=A_{c,5}T^{5}$ in a conventional three-dimensional metal. This possibility was raised by Gurzhi \cite{Gurzhi1968} and has been explored more recently by other authors \cite{Levchenko2020,Huang2021}.

A scenario which deserves special consideration here is when $\gamma_{c}$ is dominated by temperature-independent elastic scattering, because it represents an exception to the rule that the viscous regime occurs in an intermediate temperature window. The possibility that $\gamma_{c}$ can be dominated by elastic scattering has recently been put forward in the context of small-angle boundary scattering in flakes of WTe$_2$ \cite{Aharon-Steinberg2022,Wolf2023}. Suppose that $\gamma_{c}=A_{c,0}$ and that $A_{c,0}\gg A_{r,0}$. In this case the ballistic-viscous crossover is not a function of temperature, but is instead defined by $W=\lambda_{c,0}=v_{F}/\gamma_{c,0}$. This means that only one temperature-dependent crossover will occur: at high-temperature, the sample will be in an ohmic regime; at low temperature, the sample will either enter a viscous regime if its width satisfies $W\ll\lambda_{c,0}$ or a ballistic regime for $W\gg\lambda_{c,0}$. In the former case, the viscous regime would have no lower temperature limit.

The discussion up until this point has focused on the temperature-dependent occurrence of the different regimes, defining those regimes by the hierarchy of length/frequency scales. However, as discussed in \cref{sec:conductivity_results}, the behavior of the conductivity of a metal with anisotropic Fermi surface does not always match that associated with the regime as identified by the hierarchy of scales. To predict the full temperature dependent behavior in these cases, expressions for the temperature dependences of $\gamma_{r}$ and $\gamma_{c}$ can be inserted into the conductivities given in \cref{sec:conductivity_results}.

\section{Using $Z(\omega)$ to measure $\sigma(q,0)$}\label{sec:impedance}

Surface impedance for specular boundary scattering is given by \cite{Reuter1948}
\begin{equation}\label{eq:impedance_specular}
    Z_{i}^{s}
    =i\mu_{0}\omega\frac{2}{\pi}
    \int_{0}^{\infty}dq\,\mathcal{A}_{ii}(q,\omega)
\end{equation}
and for diffuse boundary scattering by \cite{Dingle1953}
\begin{equation}\label{eq:impedance_diffuse}
    Z_{i}^{d}
    =i\mu_{0}\omega\left[
        \int_{0}^{\infty}dq\,\log\left(\frac{1}{q^{2}\mathcal{A}_{ii}(q,\omega)}\right)
    \right]^{-1}
\end{equation}
where $\mathcal{A}_{ii}(q,\omega)$ is the photon propagator
\begin{equation}
    \mathcal{A}_{ii}(q,\omega)
    =\frac{1}{i\mu_{0}\omega\sigma_{ii}(q,\omega)+\omega^{2}/c^{2}-q^{2}} .
\end{equation}
For $\omega\ll\gamma_{r}$, $Z_{i}[\omega,\sigma_{ii}(q,\omega)]\approx Z_{i}[\omega,\sigma_{ii}(q,0)]$. In this case the relevant transport regime can be determined as per \cref{tab:regimes}, taking $q$ to be 
\begin{equation}
    q^{*}
    \equiv\left(\frac{\omega\gamma_{c}}{\lambda_{L}^{2}v_{F}^{2}}\right)^{1/4}
\end{equation}
where
\begin{equation}
    \lambda_{L}
    \equiv\frac{1}{\sqrt{\mu_{0}e^{2}{\langle}v_{i}|v_{i}{\rangle}}} .
\end{equation}
If the asymptotic behavior of the conductivity in this regime follows 
\begin{equation}\label{eq:sigma_asymptotic}
    \sigma_{ii}(q,0)\sim q^{-\alpha}
\end{equation}
then the asymptotic behavior of the surface impedance follows 
\begin{equation}\label{eq:z_asymptotic}
    Z_{i}\sim\omega^{\eta}\exp[-i(\pi/2)\eta]
\end{equation}
with 
\begin{equation}\label{eq:eta_asymptotic}
    \eta\equiv\frac{1+\alpha}{2+\alpha} .
\end{equation}
\Cref{eq:z_asymptotic,eq:eta_asymptotic} follow directly from substituting \cref{eq:sigma_asymptotic} into \cref{eq:impedance_specular} or \cref{eq:impedance_diffuse}.

\section{Generalization of boundary condition for channel geometry}\label{sec:BCs_general}

We define $f_{\bm{k}}^{+(-)}$ as the distribution function for electrons with $v_{\bm{k}x}>0$ ($v_{\bm{k}x}<0$). We assume that the distribution function at the boundaries of the channel follows
\begin{equation}
    f_{0}+\delta\!f_{\bm{k}}^{+}(-W/2)
    =p[f_{0}+\delta\!f_{\bm{k}}^{-}(-W/2)]+(1-p)f_{0}
\end{equation}
and 
\begin{equation}
    f_{0}+\delta\!f_{\bm{k}}^{-}(+W/2)
    =p[f_{0}+\delta\!f_{\bm{k}}^{+}(+W/2)]+(1-p)f_{0} . 
\end{equation}
This follows the common treatment applied to isotropic metals \cite{Sondheimer2001}, where $p\in[0,1]$ is interpreted as a specularity parameter with $p=1$ corresponding to completely specular boundary scattering and $p=0$ to completely diffuse boundary scattering.
Note that for an anisotropic Fermi surface, mirror symmetry in the channel is required to ensure that a specular boundary scattering event is possible. 
In this case we find that 
\begin{equation}\label{eq:M_mixedBC}
    M(x)
    =\frac{1}{\gamma_{c}}
    \left[
        1-m(p)\exp\left(
            -\frac{x}{v_{x}/\gamma_{c}}
            -\frac{W/2}{|v_{x}|/\gamma_{c}}
        \right)
    \right]
\end{equation}
with 
\begin{equation}
    m(p)\equiv\frac{1-p}{1-p\exp(-W\gamma_{c}/|v_{x}|)}
\end{equation}
We see that for $p=0$, $m(p)=1$ and we recover \cref{eq:Mx}. For $p=1$, $m(p)=0$ and $M(x)=1/\gamma_{c}$. 





\bibliography{biblio.bib}

\begin{thebibliography}{55}%
\makeatletter
\providecommand \@ifxundefined [1]{%
 \@ifx{#1\undefined}
}%
\providecommand \@ifnum [1]{%
 \ifnum #1\expandafter \@firstoftwo
 \else \expandafter \@secondoftwo
 \fi
}%
\providecommand \@ifx [1]{%
 \ifx #1\expandafter \@firstoftwo
 \else \expandafter \@secondoftwo
 \fi
}%
\providecommand \natexlab [1]{#1}%
\providecommand \enquote  [1]{``#1''}%
\providecommand \bibnamefont  [1]{#1}%
\providecommand \bibfnamefont [1]{#1}%
\providecommand \citenamefont [1]{#1}%
\providecommand \href@noop [0]{\@secondoftwo}%
\providecommand \href [0]{\begingroup \@sanitize@url \@href}%
\providecommand \@href[1]{\@@startlink{#1}\@@href}%
\providecommand \@@href[1]{\endgroup#1\@@endlink}%
\providecommand \@sanitize@url [0]{\catcode `\\12\catcode `\$12\catcode `\&12\catcode `\#12\catcode `\^12\catcode `\_12\catcode `\%12\relax}%
\providecommand \@@startlink[1]{}%
\providecommand \@@endlink[0]{}%
\providecommand \url  [0]{\begingroup\@sanitize@url \@url }%
\providecommand \@url [1]{\endgroup\@href {#1}{\urlprefix }}%
\providecommand \urlprefix  [0]{URL }%
\providecommand \Eprint [0]{\href }%
\providecommand \doibase [0]{https://doi.org/}%
\providecommand \selectlanguage [0]{\@gobble}%
\providecommand \bibinfo  [0]{\@secondoftwo}%
\providecommand \bibfield  [0]{\@secondoftwo}%
\providecommand \translation [1]{[#1]}%
\providecommand \BibitemOpen [0]{}%
\providecommand \bibitemStop [0]{}%
\providecommand \bibitemNoStop [0]{.\EOS\space}%
\providecommand \EOS [0]{\spacefactor3000\relax}%
\providecommand \BibitemShut  [1]{\csname bibitem#1\endcsname}%
\let\auto@bib@innerbib\@empty
\bibitem [{\citenamefont {Gurzhi}(1963)}]{Gurzhi1963}%
  \BibitemOpen
  \bibfield  {author} {\bibinfo {author} {\bibfnamefont {R.~N.}\ \bibnamefont {Gurzhi}},\ }\bibfield  {title} {\bibinfo {title} {{Minimum of resistance in impurity-free conductors}},\ }\href {http://jetp.ras.ru/cgi-bin/dn/e_017_02_0521.pdf} {\bibfield  {journal} {\bibinfo  {journal} {Sov. Phys. JETP}\ }\textbf {\bibinfo {volume} {44}},\ \bibinfo {pages} {771} (\bibinfo {year} {1963})}\BibitemShut {NoStop}%
\bibitem [{\citenamefont {Gurzhi}(1968)}]{Gurzhi1968}%
  \BibitemOpen
  \bibfield  {author} {\bibinfo {author} {\bibfnamefont {R.~N.}\ \bibnamefont {Gurzhi}},\ }\bibfield  {title} {\bibinfo {title} {{Hydrodynamic effects in solids at low temperature}},\ }\href {https://doi.org/10.1070/PU1968v011n02ABEH003815} {\bibfield  {journal} {\bibinfo  {journal} {Soviet Physics Uspekhi}\ }\textbf {\bibinfo {volume} {11}},\ \bibinfo {pages} {255} (\bibinfo {year} {1968})}\BibitemShut {NoStop}%
\bibitem [{\citenamefont {Molenkamp}\ and\ \citenamefont {{De Jong}}(1994)}]{Molenkamp1994}%
  \BibitemOpen
  \bibfield  {author} {\bibinfo {author} {\bibfnamefont {L.~W.}\ \bibnamefont {Molenkamp}}\ and\ \bibinfo {author} {\bibfnamefont {M.~J.}\ \bibnamefont {{De Jong}}},\ }\bibfield  {title} {\bibinfo {title} {{Electron-electron-scattering-induced size effects in a two-dimensional wire}},\ }\href {https://doi.org/10.1103/PhysRevB.49.5038} {\bibfield  {journal} {\bibinfo  {journal} {Physical Review B}\ }\textbf {\bibinfo {volume} {49}},\ \bibinfo {pages} {5038} (\bibinfo {year} {1994})}\BibitemShut {NoStop}%
\bibitem [{\citenamefont {DeJong}\ and\ \citenamefont {Molenkamp}(1995)}]{DeJong1995}%
  \BibitemOpen
  \bibfield  {author} {\bibinfo {author} {\bibfnamefont {M.~J.~M.}\ \bibnamefont {DeJong}}\ and\ \bibinfo {author} {\bibfnamefont {L.~W.}\ \bibnamefont {Molenkamp}},\ }\bibfield  {title} {\bibinfo {title} {{Hydrodynamic electron flow in high-mobility wires}},\ }\href {https://doi.org/10.1103/PhysRevB.51.13389} {\bibfield  {journal} {\bibinfo  {journal} {Physical Review B}\ }\textbf {\bibinfo {volume} {51}},\ \bibinfo {pages} {13389} (\bibinfo {year} {1995})}\BibitemShut {NoStop}%
\bibitem [{\citenamefont {Crossno}\ \emph {et~al.}(2016)\citenamefont {Crossno}, \citenamefont {Shi}, \citenamefont {Wang}, \citenamefont {Liu}, \citenamefont {Harzheim}, \citenamefont {Lucas}, \citenamefont {Sachdev}, \citenamefont {Kim}, \citenamefont {Taniguchi}, \citenamefont {Watanabe}, \citenamefont {Ohki},\ and\ \citenamefont {Fong}}]{Crossno2016}%
  \BibitemOpen
  \bibfield  {author} {\bibinfo {author} {\bibfnamefont {J.}~\bibnamefont {Crossno}}, \bibinfo {author} {\bibfnamefont {J.~K.}\ \bibnamefont {Shi}}, \bibinfo {author} {\bibfnamefont {K.}~\bibnamefont {Wang}}, \bibinfo {author} {\bibfnamefont {X.}~\bibnamefont {Liu}}, \bibinfo {author} {\bibfnamefont {A.}~\bibnamefont {Harzheim}}, \bibinfo {author} {\bibfnamefont {A.}~\bibnamefont {Lucas}}, \bibinfo {author} {\bibfnamefont {S.}~\bibnamefont {Sachdev}}, \bibinfo {author} {\bibfnamefont {P.}~\bibnamefont {Kim}}, \bibinfo {author} {\bibfnamefont {T.}~\bibnamefont {Taniguchi}}, \bibinfo {author} {\bibfnamefont {K.}~\bibnamefont {Watanabe}}, \bibinfo {author} {\bibfnamefont {T.~A.}\ \bibnamefont {Ohki}},\ and\ \bibinfo {author} {\bibfnamefont {K.~C.}\ \bibnamefont {Fong}},\ }\bibfield  {title} {\bibinfo {title} {{Observation of the Dirac fluid and the breakdown of the Wiedemann-Franz law in graphene}},\ }\href {https://doi.org/10.1126/science.aad0343} {\bibfield  {journal} {\bibinfo  {journal} {Science}\ }\textbf
  {\bibinfo {volume} {351}},\ \bibinfo {pages} {1058} (\bibinfo {year} {2016})}\BibitemShut {NoStop}%
\bibitem [{\citenamefont {Bandurin}\ \emph {et~al.}(2016)\citenamefont {Bandurin}, \citenamefont {Torre}, \citenamefont {{Krishna Kumar}}, \citenamefont {{Ben Shalom}}, \citenamefont {Tomadin}, \citenamefont {Principi}, \citenamefont {Auton}, \citenamefont {Khestanova}, \citenamefont {Novoselov}, \citenamefont {Grigorieva}, \citenamefont {Ponomarenko}, \citenamefont {Geim},\ and\ \citenamefont {Polini}}]{Bandurin2016}%
  \BibitemOpen
  \bibfield  {author} {\bibinfo {author} {\bibfnamefont {D.~A.}\ \bibnamefont {Bandurin}}, \bibinfo {author} {\bibfnamefont {I.}~\bibnamefont {Torre}}, \bibinfo {author} {\bibfnamefont {R.}~\bibnamefont {{Krishna Kumar}}}, \bibinfo {author} {\bibfnamefont {M.}~\bibnamefont {{Ben Shalom}}}, \bibinfo {author} {\bibfnamefont {A.}~\bibnamefont {Tomadin}}, \bibinfo {author} {\bibfnamefont {A.}~\bibnamefont {Principi}}, \bibinfo {author} {\bibfnamefont {G.~H.}\ \bibnamefont {Auton}}, \bibinfo {author} {\bibfnamefont {E.}~\bibnamefont {Khestanova}}, \bibinfo {author} {\bibfnamefont {K.~S.}\ \bibnamefont {Novoselov}}, \bibinfo {author} {\bibfnamefont {I.~V.}\ \bibnamefont {Grigorieva}}, \bibinfo {author} {\bibfnamefont {L.~A.}\ \bibnamefont {Ponomarenko}}, \bibinfo {author} {\bibfnamefont {A.~K.}\ \bibnamefont {Geim}},\ and\ \bibinfo {author} {\bibfnamefont {M.}~\bibnamefont {Polini}},\ }\bibfield  {title} {\bibinfo {title} {{Negative local resistance caused by viscous electron backflow in graphene}},\ }\href
  {https://doi.org/10.1126/science.aad0201} {\bibfield  {journal} {\bibinfo  {journal} {Science}\ }\textbf {\bibinfo {volume} {351}},\ \bibinfo {pages} {1055} (\bibinfo {year} {2016})}\BibitemShut {NoStop}%
\bibitem [{\citenamefont {{Krishna Kumar}}\ \emph {et~al.}(2017)\citenamefont {{Krishna Kumar}}, \citenamefont {Bandurin}, \citenamefont {{D Pellegrino}}, \citenamefont {Cao}, \citenamefont {Principi}, \citenamefont {Guo}, \citenamefont {Auton}, \citenamefont {{Ben Shalom}}, \citenamefont {Ponomarenko}, \citenamefont {Falkovich}, \citenamefont {Watanabe}, \citenamefont {Taniguchi}, \citenamefont {Grigorieva}, \citenamefont {Levitov}, \citenamefont {Polini},\ and\ \citenamefont {Geim}}]{KrishnaKumar2017}%
  \BibitemOpen
  \bibfield  {author} {\bibinfo {author} {\bibfnamefont {R.}~\bibnamefont {{Krishna Kumar}}}, \bibinfo {author} {\bibfnamefont {D.~A.}\ \bibnamefont {Bandurin}}, \bibinfo {author} {\bibfnamefont {F.~M.}\ \bibnamefont {{D Pellegrino}}}, \bibinfo {author} {\bibfnamefont {Y.}~\bibnamefont {Cao}}, \bibinfo {author} {\bibfnamefont {A.}~\bibnamefont {Principi}}, \bibinfo {author} {\bibfnamefont {H.}~\bibnamefont {Guo}}, \bibinfo {author} {\bibfnamefont {G.~H.}\ \bibnamefont {Auton}}, \bibinfo {author} {\bibfnamefont {M.}~\bibnamefont {{Ben Shalom}}}, \bibinfo {author} {\bibfnamefont {L.~A.}\ \bibnamefont {Ponomarenko}}, \bibinfo {author} {\bibfnamefont {G.}~\bibnamefont {Falkovich}}, \bibinfo {author} {\bibfnamefont {K.}~\bibnamefont {Watanabe}}, \bibinfo {author} {\bibfnamefont {T.}~\bibnamefont {Taniguchi}}, \bibinfo {author} {\bibfnamefont {I.~V.}\ \bibnamefont {Grigorieva}}, \bibinfo {author} {\bibfnamefont {L.~S.}\ \bibnamefont {Levitov}}, \bibinfo {author} {\bibfnamefont {M.}~\bibnamefont {Polini}},\ and\
  \bibinfo {author} {\bibfnamefont {A.~K.}\ \bibnamefont {Geim}},\ }\bibfield  {title} {\bibinfo {title} {{Superballistic flow of viscous electron fluid through graphene constrictions}},\ }\href {https://doi.org/10.1038/NPHYS4240} {\bibfield  {journal} {\bibinfo  {journal} {Nature Physics}\ }\textbf {\bibinfo {volume} {13}},\ \bibinfo {pages} {1182} (\bibinfo {year} {2017})}\BibitemShut {NoStop}%
\bibitem [{\citenamefont {Sulpizio}\ \emph {et~al.}(2019)\citenamefont {Sulpizio}, \citenamefont {Ella}, \citenamefont {Rozen}, \citenamefont {Birkbeck}, \citenamefont {Perello}, \citenamefont {Dutta}, \citenamefont {Ben-Shalom}, \citenamefont {Taniguchi}, \citenamefont {Watanabe}, \citenamefont {Holder}, \citenamefont {Queiroz}, \citenamefont {Principi}, \citenamefont {Stern}, \citenamefont {Scaffidi}, \citenamefont {Geim},\ and\ \citenamefont {Ilani}}]{Sulpizio2019}%
  \BibitemOpen
  \bibfield  {author} {\bibinfo {author} {\bibfnamefont {J.~A.}\ \bibnamefont {Sulpizio}}, \bibinfo {author} {\bibfnamefont {L.}~\bibnamefont {Ella}}, \bibinfo {author} {\bibfnamefont {A.}~\bibnamefont {Rozen}}, \bibinfo {author} {\bibfnamefont {J.}~\bibnamefont {Birkbeck}}, \bibinfo {author} {\bibfnamefont {D.~J.}\ \bibnamefont {Perello}}, \bibinfo {author} {\bibfnamefont {D.}~\bibnamefont {Dutta}}, \bibinfo {author} {\bibfnamefont {M.}~\bibnamefont {Ben-Shalom}}, \bibinfo {author} {\bibfnamefont {T.}~\bibnamefont {Taniguchi}}, \bibinfo {author} {\bibfnamefont {K.}~\bibnamefont {Watanabe}}, \bibinfo {author} {\bibfnamefont {T.}~\bibnamefont {Holder}}, \bibinfo {author} {\bibfnamefont {R.}~\bibnamefont {Queiroz}}, \bibinfo {author} {\bibfnamefont {A.}~\bibnamefont {Principi}}, \bibinfo {author} {\bibfnamefont {A.}~\bibnamefont {Stern}}, \bibinfo {author} {\bibfnamefont {T.}~\bibnamefont {Scaffidi}}, \bibinfo {author} {\bibfnamefont {A.~K.}\ \bibnamefont {Geim}},\ and\ \bibinfo {author} {\bibfnamefont
  {S.}~\bibnamefont {Ilani}},\ }\bibfield  {title} {\bibinfo {title} {{Visualizing Poiseuille flow of hydrodynamic electrons}},\ }\href {https://doi.org/10.1038/s41586-019-1788-9} {\bibfield  {journal} {\bibinfo  {journal} {Nature}\ }\textbf {\bibinfo {volume} {576}},\ \bibinfo {pages} {75} (\bibinfo {year} {2019})}\BibitemShut {NoStop}%
\bibitem [{\citenamefont {Ku}\ \emph {et~al.}(2020)\citenamefont {Ku}, \citenamefont {Zhou}, \citenamefont {Li}, \citenamefont {Shin}, \citenamefont {Shi}, \citenamefont {Burch}, \citenamefont {Anderson}, \citenamefont {Pierce}, \citenamefont {Xie}, \citenamefont {Hamo}, \citenamefont {Vool}, \citenamefont {Zhang}, \citenamefont {Casola}, \citenamefont {Taniguchi}, \citenamefont {Watanabe}, \citenamefont {Fogler}, \citenamefont {Kim}, \citenamefont {Yacoby},\ and\ \citenamefont {Walsworth}}]{Ku2020}%
  \BibitemOpen
  \bibfield  {author} {\bibinfo {author} {\bibfnamefont {M.~J.}\ \bibnamefont {Ku}}, \bibinfo {author} {\bibfnamefont {T.~X.}\ \bibnamefont {Zhou}}, \bibinfo {author} {\bibfnamefont {Q.}~\bibnamefont {Li}}, \bibinfo {author} {\bibfnamefont {Y.~J.}\ \bibnamefont {Shin}}, \bibinfo {author} {\bibfnamefont {J.~K.}\ \bibnamefont {Shi}}, \bibinfo {author} {\bibfnamefont {C.}~\bibnamefont {Burch}}, \bibinfo {author} {\bibfnamefont {L.~E.}\ \bibnamefont {Anderson}}, \bibinfo {author} {\bibfnamefont {A.~T.}\ \bibnamefont {Pierce}}, \bibinfo {author} {\bibfnamefont {Y.}~\bibnamefont {Xie}}, \bibinfo {author} {\bibfnamefont {A.}~\bibnamefont {Hamo}}, \bibinfo {author} {\bibfnamefont {U.}~\bibnamefont {Vool}}, \bibinfo {author} {\bibfnamefont {H.}~\bibnamefont {Zhang}}, \bibinfo {author} {\bibfnamefont {F.}~\bibnamefont {Casola}}, \bibinfo {author} {\bibfnamefont {T.}~\bibnamefont {Taniguchi}}, \bibinfo {author} {\bibfnamefont {K.}~\bibnamefont {Watanabe}}, \bibinfo {author} {\bibfnamefont {M.~M.}\ \bibnamefont
  {Fogler}}, \bibinfo {author} {\bibfnamefont {P.}~\bibnamefont {Kim}}, \bibinfo {author} {\bibfnamefont {A.}~\bibnamefont {Yacoby}},\ and\ \bibinfo {author} {\bibfnamefont {R.~L.}\ \bibnamefont {Walsworth}},\ }\bibfield  {title} {\bibinfo {title} {{Imaging viscous flow of the Dirac fluid in graphene}},\ }\href {https://doi.org/10.1038/s41586-020-2507-2} {\bibfield  {journal} {\bibinfo  {journal} {Nature}\ }\textbf {\bibinfo {volume} {583}},\ \bibinfo {pages} {537} (\bibinfo {year} {2020})},\ \Eprint {https://arxiv.org/abs/1905.10791} {arXiv:1905.10791} \BibitemShut {NoStop}%
\bibitem [{\citenamefont {Kumar}\ \emph {et~al.}(2022)\citenamefont {Kumar}, \citenamefont {Birkbeck}, \citenamefont {Sulpizio}, \citenamefont {Perello}, \citenamefont {Taniguchi}, \citenamefont {Watanabe}, \citenamefont {Reuven}, \citenamefont {Scaffidi}, \citenamefont {Stern}, \citenamefont {Geim},\ and\ \citenamefont {Ilani}}]{Kumar2022}%
  \BibitemOpen
  \bibfield  {author} {\bibinfo {author} {\bibfnamefont {C.}~\bibnamefont {Kumar}}, \bibinfo {author} {\bibfnamefont {J.}~\bibnamefont {Birkbeck}}, \bibinfo {author} {\bibfnamefont {J.~A.}\ \bibnamefont {Sulpizio}}, \bibinfo {author} {\bibfnamefont {D.}~\bibnamefont {Perello}}, \bibinfo {author} {\bibfnamefont {T.}~\bibnamefont {Taniguchi}}, \bibinfo {author} {\bibfnamefont {K.}~\bibnamefont {Watanabe}}, \bibinfo {author} {\bibfnamefont {O.}~\bibnamefont {Reuven}}, \bibinfo {author} {\bibfnamefont {T.}~\bibnamefont {Scaffidi}}, \bibinfo {author} {\bibfnamefont {A.}~\bibnamefont {Stern}}, \bibinfo {author} {\bibfnamefont {A.~K.}\ \bibnamefont {Geim}},\ and\ \bibinfo {author} {\bibfnamefont {S.}~\bibnamefont {Ilani}},\ }\bibfield  {title} {\bibinfo {title} {{Imaging hydrodynamic electrons flowing without Landauer–Sharvin resistance}},\ }\href {https://doi.org/10.1038/s41586-022-05002-7} {\bibfield  {journal} {\bibinfo  {journal} {Nature}\ }\textbf {\bibinfo {volume} {609}},\ \bibinfo {pages} {276} (\bibinfo
  {year} {2022})}\BibitemShut {NoStop}%
\bibitem [{\citenamefont {Moll}\ \emph {et~al.}(2016)\citenamefont {Moll}, \citenamefont {Kushwaha}, \citenamefont {Nandi}, \citenamefont {Schmidt},\ and\ \citenamefont {Mackenzie}}]{Moll2016}%
  \BibitemOpen
  \bibfield  {author} {\bibinfo {author} {\bibfnamefont {P.~J.~W.}\ \bibnamefont {Moll}}, \bibinfo {author} {\bibfnamefont {P.}~\bibnamefont {Kushwaha}}, \bibinfo {author} {\bibfnamefont {N.}~\bibnamefont {Nandi}}, \bibinfo {author} {\bibfnamefont {B.}~\bibnamefont {Schmidt}},\ and\ \bibinfo {author} {\bibfnamefont {A.~P.}\ \bibnamefont {Mackenzie}},\ }\bibfield  {title} {\bibinfo {title} {{Evidence for hydrodynamic electron flow in PdCoO$_2$}},\ }\href {https://doi.org/10.1126/science.aac8385} {\bibfield  {journal} {\bibinfo  {journal} {Science}\ }\textbf {\bibinfo {volume} {351}},\ \bibinfo {pages} {1061} (\bibinfo {year} {2016})}\BibitemShut {NoStop}%
\bibitem [{\citenamefont {Bachmann}\ \emph {et~al.}(2019)\citenamefont {Bachmann}, \citenamefont {Sharpe}, \citenamefont {Barnard}, \citenamefont {Putzke}, \citenamefont {K{\"{o}}nig}, \citenamefont {Khim}, \citenamefont {Goldhaber-Gordon}, \citenamefont {Mackenzie},\ and\ \citenamefont {Moll}}]{Bachmann2019}%
  \BibitemOpen
  \bibfield  {author} {\bibinfo {author} {\bibfnamefont {M.~D.}\ \bibnamefont {Bachmann}}, \bibinfo {author} {\bibfnamefont {A.~L.}\ \bibnamefont {Sharpe}}, \bibinfo {author} {\bibfnamefont {A.~W.}\ \bibnamefont {Barnard}}, \bibinfo {author} {\bibfnamefont {C.}~\bibnamefont {Putzke}}, \bibinfo {author} {\bibfnamefont {M.}~\bibnamefont {K{\"{o}}nig}}, \bibinfo {author} {\bibfnamefont {S.}~\bibnamefont {Khim}}, \bibinfo {author} {\bibfnamefont {D.}~\bibnamefont {Goldhaber-Gordon}}, \bibinfo {author} {\bibfnamefont {A.~P.}\ \bibnamefont {Mackenzie}},\ and\ \bibinfo {author} {\bibfnamefont {P.~J.~W.}\ \bibnamefont {Moll}},\ }\bibfield  {title} {\bibinfo {title} {{Super-geometric electron focusing on the hexagonal Fermi surface of PdCoO$_2$}},\ }\href {https://doi.org/10.1038/s41467-019-13020-9} {\bibfield  {journal} {\bibinfo  {journal} {Nature Communications}\ }\textbf {\bibinfo {volume} {10}},\ \bibinfo {pages} {5081} (\bibinfo {year} {2019})}\BibitemShut {NoStop}%
\bibitem [{\citenamefont {McGuinness}\ \emph {et~al.}(2021)\citenamefont {McGuinness}, \citenamefont {Zhakina}, \citenamefont {K{\"{o}}nig}, \citenamefont {Bachmann}, \citenamefont {Putzke}, \citenamefont {Moll}, \citenamefont {Khim},\ and\ \citenamefont {Mackenzie}}]{McGuinness2021}%
  \BibitemOpen
  \bibfield  {author} {\bibinfo {author} {\bibfnamefont {P.~H.}\ \bibnamefont {McGuinness}}, \bibinfo {author} {\bibfnamefont {E.}~\bibnamefont {Zhakina}}, \bibinfo {author} {\bibfnamefont {M.}~\bibnamefont {K{\"{o}}nig}}, \bibinfo {author} {\bibfnamefont {M.~D.}\ \bibnamefont {Bachmann}}, \bibinfo {author} {\bibfnamefont {C.}~\bibnamefont {Putzke}}, \bibinfo {author} {\bibfnamefont {P.~J.~W.}\ \bibnamefont {Moll}}, \bibinfo {author} {\bibfnamefont {S.}~\bibnamefont {Khim}},\ and\ \bibinfo {author} {\bibfnamefont {A.~P.}\ \bibnamefont {Mackenzie}},\ }\bibfield  {title} {\bibinfo {title} {{Low-symmetry nonlocal transport in microstructured squares of delafossite metals}},\ }\href {https://doi.org/10.1073/PNAS.2113185118} {\bibfield  {journal} {\bibinfo  {journal} {Proceedings of the National Academy of Sciences}\ }\textbf {\bibinfo {volume} {118}},\ \bibinfo {pages} {2113185118} (\bibinfo {year} {2021})}\BibitemShut {NoStop}%
\bibitem [{\citenamefont {Bachmann}\ \emph {et~al.}(2022)\citenamefont {Bachmann}, \citenamefont {Sharpe}, \citenamefont {Baker}, \citenamefont {Barnard}, \citenamefont {Putzke}, \citenamefont {Scaffidi}, \citenamefont {Nandi}, \citenamefont {Mcguinness}, \citenamefont {Zhakina}, \citenamefont {Moravec}, \citenamefont {Khim}, \citenamefont {K{\"{o}}nig}, \citenamefont {Goldhaber-Gordon}, \citenamefont {Mackenzie},\ and\ \citenamefont {Moll}}]{Bachmann2022}%
  \BibitemOpen
  \bibfield  {author} {\bibinfo {author} {\bibfnamefont {M.~D.}\ \bibnamefont {Bachmann}}, \bibinfo {author} {\bibfnamefont {A.~L.}\ \bibnamefont {Sharpe}}, \bibinfo {author} {\bibfnamefont {G.}~\bibnamefont {Baker}}, \bibinfo {author} {\bibfnamefont {A.~W.}\ \bibnamefont {Barnard}}, \bibinfo {author} {\bibfnamefont {C.}~\bibnamefont {Putzke}}, \bibinfo {author} {\bibfnamefont {T.}~\bibnamefont {Scaffidi}}, \bibinfo {author} {\bibfnamefont {N.}~\bibnamefont {Nandi}}, \bibinfo {author} {\bibfnamefont {P.~H.}\ \bibnamefont {Mcguinness}}, \bibinfo {author} {\bibfnamefont {E.}~\bibnamefont {Zhakina}}, \bibinfo {author} {\bibfnamefont {M.}~\bibnamefont {Moravec}}, \bibinfo {author} {\bibfnamefont {S.}~\bibnamefont {Khim}}, \bibinfo {author} {\bibfnamefont {M.}~\bibnamefont {K{\"{o}}nig}}, \bibinfo {author} {\bibfnamefont {D.}~\bibnamefont {Goldhaber-Gordon}}, \bibinfo {author} {\bibfnamefont {A.~P.}\ \bibnamefont {Mackenzie}},\ and\ \bibinfo {author} {\bibfnamefont {P.~J.~W.}\ \bibnamefont {Moll}},\ }\bibfield
  {title} {\bibinfo {title} {{Directional ballistic transport in the two-dimensional metal PdCoO$_2$}},\ }\href {https://doi.org/10.1038/s41567-022-01570-7} {\bibfield  {journal} {\bibinfo  {journal} {Nature Physics}\ }\textbf {\bibinfo {volume} {18}},\ \bibinfo {pages} {819} (\bibinfo {year} {2022})}\BibitemShut {NoStop}%
\bibitem [{\citenamefont {Baker}\ \emph {et~al.}(2022)\citenamefont {Baker}, \citenamefont {Branch}, \citenamefont {Day}, \citenamefont {Valentinis}, \citenamefont {Oudah}, \citenamefont {McGuinness}, \citenamefont {Khim}, \citenamefont {Sur{\'{o}}wka}, \citenamefont {Moessner}, \citenamefont {Schmalian}, \citenamefont {Mackenzie},\ and\ \citenamefont {Bonn}}]{Baker2022}%
  \BibitemOpen
  \bibfield  {author} {\bibinfo {author} {\bibfnamefont {G.}~\bibnamefont {Baker}}, \bibinfo {author} {\bibfnamefont {T.~W.}\ \bibnamefont {Branch}}, \bibinfo {author} {\bibfnamefont {J.}~\bibnamefont {Day}}, \bibinfo {author} {\bibfnamefont {D.}~\bibnamefont {Valentinis}}, \bibinfo {author} {\bibfnamefont {M.}~\bibnamefont {Oudah}}, \bibinfo {author} {\bibfnamefont {P.}~\bibnamefont {McGuinness}}, \bibinfo {author} {\bibfnamefont {S.}~\bibnamefont {Khim}}, \bibinfo {author} {\bibfnamefont {P.}~\bibnamefont {Sur{\'{o}}wka}}, \bibinfo {author} {\bibfnamefont {R.}~\bibnamefont {Moessner}}, \bibinfo {author} {\bibfnamefont {J.}~\bibnamefont {Schmalian}}, \bibinfo {author} {\bibfnamefont {A.~P.}\ \bibnamefont {Mackenzie}},\ and\ \bibinfo {author} {\bibfnamefont {D.~A.}\ \bibnamefont {Bonn}},\ }\bibfield  {title} {\bibinfo {title} {{Non-local microwave electrodynamics in ultra-pure PdCoO$_2$}},\ }\bibfield  {journal} {\bibinfo  {journal} {arXiv}\ }\href {https://doi.org/10.48550/arXiv.2204.14239}
  {10.48550/arXiv.2204.14239} (\bibinfo {year} {2022})\BibitemShut {NoStop}%
\bibitem [{\citenamefont {Zhakina}\ \emph {et~al.}(2023)\citenamefont {Zhakina}, \citenamefont {McGuinness}, \citenamefont {K{\"{o}}nig}, \citenamefont {Grasset}, \citenamefont {Bachmann}, \citenamefont {Khim}, \citenamefont {Putzke}, \citenamefont {Moll}, \citenamefont {Konczykowski},\ and\ \citenamefont {Mackenzie}}]{Zhakina2023}%
  \BibitemOpen
  \bibfield  {author} {\bibinfo {author} {\bibfnamefont {E.}~\bibnamefont {Zhakina}}, \bibinfo {author} {\bibfnamefont {P.~H.}\ \bibnamefont {McGuinness}}, \bibinfo {author} {\bibfnamefont {M.}~\bibnamefont {K{\"{o}}nig}}, \bibinfo {author} {\bibfnamefont {R.}~\bibnamefont {Grasset}}, \bibinfo {author} {\bibfnamefont {M.~D.}\ \bibnamefont {Bachmann}}, \bibinfo {author} {\bibfnamefont {S.}~\bibnamefont {Khim}}, \bibinfo {author} {\bibfnamefont {C.}~\bibnamefont {Putzke}}, \bibinfo {author} {\bibfnamefont {P.~J.~W.}\ \bibnamefont {Moll}}, \bibinfo {author} {\bibfnamefont {M.}~\bibnamefont {Konczykowski}},\ and\ \bibinfo {author} {\bibfnamefont {A.~P.}\ \bibnamefont {Mackenzie}},\ }\bibfield  {title} {\bibinfo {title} {{Crossing the ballistic-ohmic transition via high energy electron irradiation}},\ }\href {https://doi.org/10.1103/physrevb.107.094203} {\bibfield  {journal} {\bibinfo  {journal} {Physical Review B}\ }\textbf {\bibinfo {volume} {107}},\ \bibinfo {pages} {094203} (\bibinfo {year}
  {2023})}\BibitemShut {NoStop}%
\bibitem [{\citenamefont {Ali}\ \emph {et~al.}(2014)\citenamefont {Ali}, \citenamefont {Xiong}, \citenamefont {Flynn}, \citenamefont {Tao}, \citenamefont {Gibson}, \citenamefont {Schoop}, \citenamefont {Liang}, \citenamefont {Haldolaarachchige}, \citenamefont {Hirschberger}, \citenamefont {Ong},\ and\ \citenamefont {Cava}}]{Ali2014}%
  \BibitemOpen
  \bibfield  {author} {\bibinfo {author} {\bibfnamefont {M.~N.}\ \bibnamefont {Ali}}, \bibinfo {author} {\bibfnamefont {J.}~\bibnamefont {Xiong}}, \bibinfo {author} {\bibfnamefont {S.}~\bibnamefont {Flynn}}, \bibinfo {author} {\bibfnamefont {J.}~\bibnamefont {Tao}}, \bibinfo {author} {\bibfnamefont {Q.~D.}\ \bibnamefont {Gibson}}, \bibinfo {author} {\bibfnamefont {L.~M.}\ \bibnamefont {Schoop}}, \bibinfo {author} {\bibfnamefont {T.}~\bibnamefont {Liang}}, \bibinfo {author} {\bibfnamefont {N.}~\bibnamefont {Haldolaarachchige}}, \bibinfo {author} {\bibfnamefont {M.}~\bibnamefont {Hirschberger}}, \bibinfo {author} {\bibfnamefont {N.~P.}\ \bibnamefont {Ong}},\ and\ \bibinfo {author} {\bibfnamefont {R.~J.}\ \bibnamefont {Cava}},\ }\bibfield  {title} {\bibinfo {title} {{Large, non-saturating magnetoresistance in WTe$_2$}},\ }\href {https://doi.org/10.1038/nature13763} {\bibfield  {journal} {\bibinfo  {journal} {Nature}\ }\textbf {\bibinfo {volume} {514}},\ \bibinfo {pages} {205} (\bibinfo {year}
  {2014})}\BibitemShut {NoStop}%
\bibitem [{\citenamefont {Zhu}\ \emph {et~al.}(2015)\citenamefont {Zhu}, \citenamefont {Lin}, \citenamefont {Liu}, \citenamefont {Fauqu{\'{e}}}, \citenamefont {Tao}, \citenamefont {Yang}, \citenamefont {Shi},\ and\ \citenamefont {Behnia}}]{Zhu2015}%
  \BibitemOpen
  \bibfield  {author} {\bibinfo {author} {\bibfnamefont {Z.}~\bibnamefont {Zhu}}, \bibinfo {author} {\bibfnamefont {X.}~\bibnamefont {Lin}}, \bibinfo {author} {\bibfnamefont {J.}~\bibnamefont {Liu}}, \bibinfo {author} {\bibfnamefont {B.}~\bibnamefont {Fauqu{\'{e}}}}, \bibinfo {author} {\bibfnamefont {Q.}~\bibnamefont {Tao}}, \bibinfo {author} {\bibfnamefont {C.}~\bibnamefont {Yang}}, \bibinfo {author} {\bibfnamefont {Y.}~\bibnamefont {Shi}},\ and\ \bibinfo {author} {\bibfnamefont {K.}~\bibnamefont {Behnia}},\ }\bibfield  {title} {\bibinfo {title} {{Quantum oscillations, thermoelectric coefficients, and the fermi surface of semimetallic WTe$_2$}},\ }\href {https://doi.org/10.1103/PhysRevLett.114.176601} {\bibfield  {journal} {\bibinfo  {journal} {Physical Review Letters}\ }\textbf {\bibinfo {volume} {114}},\ \bibinfo {pages} {176601} (\bibinfo {year} {2015})}\BibitemShut {NoStop}%
\bibitem [{\citenamefont {Gooth}\ \emph {et~al.}(2018)\citenamefont {Gooth}, \citenamefont {Menges}, \citenamefont {Kumar}, \citenamefont {S{\"{u}}ss}, \citenamefont {Shekhar}, \citenamefont {Sun}, \citenamefont {Drechsler},\ and\ \citenamefont {Zierold}}]{Gooth2017}%
  \BibitemOpen
  \bibfield  {author} {\bibinfo {author} {\bibfnamefont {J.}~\bibnamefont {Gooth}}, \bibinfo {author} {\bibfnamefont {F.}~\bibnamefont {Menges}}, \bibinfo {author} {\bibfnamefont {N.}~\bibnamefont {Kumar}}, \bibinfo {author} {\bibfnamefont {V.}~\bibnamefont {S{\"{u}}ss}}, \bibinfo {author} {\bibfnamefont {C.}~\bibnamefont {Shekhar}}, \bibinfo {author} {\bibfnamefont {Y.}~\bibnamefont {Sun}}, \bibinfo {author} {\bibfnamefont {U.}~\bibnamefont {Drechsler}},\ and\ \bibinfo {author} {\bibfnamefont {R.}~\bibnamefont {Zierold}},\ }\bibfield  {title} {\bibinfo {title} {{Thermal and electrical signatures of a hydrodynamic electron fluid in WP2}},\ }\bibfield  {journal} {\bibinfo  {journal} {Nature Communications}\ }\textbf {\bibinfo {volume} {9}},\ \href {https://doi.org/10.1038/s41467-018-06688-y} {10.1038/s41467-018-06688-y} (\bibinfo {year} {2018}),\ \Eprint {https://arxiv.org/abs/1706.05925} {arXiv:1706.05925} \BibitemShut {NoStop}%
\bibitem [{\citenamefont {van Delft}\ \emph {et~al.}(2021)\citenamefont {van Delft}, \citenamefont {Wang}, \citenamefont {Putzke}, \citenamefont {Oswald}, \citenamefont {Varnavides}, \citenamefont {Garcia}, \citenamefont {Guo}, \citenamefont {Schmid}, \citenamefont {S{\"{u}}ss}, \citenamefont {Borrmann}, \citenamefont {Diaz}, \citenamefont {Sun}, \citenamefont {Felser}, \citenamefont {Gotsmann}, \citenamefont {Narang},\ and\ \citenamefont {Moll}}]{VanDelft2021}%
  \BibitemOpen
  \bibfield  {author} {\bibinfo {author} {\bibfnamefont {M.~R.}\ \bibnamefont {van Delft}}, \bibinfo {author} {\bibfnamefont {Y.}~\bibnamefont {Wang}}, \bibinfo {author} {\bibfnamefont {C.}~\bibnamefont {Putzke}}, \bibinfo {author} {\bibfnamefont {J.}~\bibnamefont {Oswald}}, \bibinfo {author} {\bibfnamefont {G.}~\bibnamefont {Varnavides}}, \bibinfo {author} {\bibfnamefont {C.~A.}\ \bibnamefont {Garcia}}, \bibinfo {author} {\bibfnamefont {C.}~\bibnamefont {Guo}}, \bibinfo {author} {\bibfnamefont {H.}~\bibnamefont {Schmid}}, \bibinfo {author} {\bibfnamefont {V.}~\bibnamefont {S{\"{u}}ss}}, \bibinfo {author} {\bibfnamefont {H.}~\bibnamefont {Borrmann}}, \bibinfo {author} {\bibfnamefont {J.}~\bibnamefont {Diaz}}, \bibinfo {author} {\bibfnamefont {Y.}~\bibnamefont {Sun}}, \bibinfo {author} {\bibfnamefont {C.}~\bibnamefont {Felser}}, \bibinfo {author} {\bibfnamefont {B.}~\bibnamefont {Gotsmann}}, \bibinfo {author} {\bibfnamefont {P.}~\bibnamefont {Narang}},\ and\ \bibinfo {author} {\bibfnamefont {P.~J.}\
  \bibnamefont {Moll}},\ }\bibfield  {title} {\bibinfo {title} {{Sondheimer oscillations as a probe of non-ohmic flow in WP2 crystals}},\ }\href {https://doi.org/10.1038/s41467-021-25037-0} {\bibfield  {journal} {\bibinfo  {journal} {Nature Communications}\ }\textbf {\bibinfo {volume} {12}},\ \bibinfo {pages} {4799} (\bibinfo {year} {2021})}\BibitemShut {NoStop}%
\bibitem [{\citenamefont {Vool}\ \emph {et~al.}(2021)\citenamefont {Vool}, \citenamefont {Hamo}, \citenamefont {Varnavides}, \citenamefont {Wang}, \citenamefont {Zhou}, \citenamefont {Kumar}, \citenamefont {Dovzhenko}, \citenamefont {Qiu}, \citenamefont {Garcia}, \citenamefont {Pierce}, \citenamefont {Gooth}, \citenamefont {Anikeeva}, \citenamefont {Felser}, \citenamefont {Narang},\ and\ \citenamefont {Yacoby}}]{Vool2021}%
  \BibitemOpen
  \bibfield  {author} {\bibinfo {author} {\bibfnamefont {U.}~\bibnamefont {Vool}}, \bibinfo {author} {\bibfnamefont {A.}~\bibnamefont {Hamo}}, \bibinfo {author} {\bibfnamefont {G.}~\bibnamefont {Varnavides}}, \bibinfo {author} {\bibfnamefont {Y.}~\bibnamefont {Wang}}, \bibinfo {author} {\bibfnamefont {T.~X.}\ \bibnamefont {Zhou}}, \bibinfo {author} {\bibfnamefont {N.}~\bibnamefont {Kumar}}, \bibinfo {author} {\bibfnamefont {Y.}~\bibnamefont {Dovzhenko}}, \bibinfo {author} {\bibfnamefont {Z.}~\bibnamefont {Qiu}}, \bibinfo {author} {\bibfnamefont {C.~A.}\ \bibnamefont {Garcia}}, \bibinfo {author} {\bibfnamefont {A.~T.}\ \bibnamefont {Pierce}}, \bibinfo {author} {\bibfnamefont {J.}~\bibnamefont {Gooth}}, \bibinfo {author} {\bibfnamefont {P.}~\bibnamefont {Anikeeva}}, \bibinfo {author} {\bibfnamefont {C.}~\bibnamefont {Felser}}, \bibinfo {author} {\bibfnamefont {P.}~\bibnamefont {Narang}},\ and\ \bibinfo {author} {\bibfnamefont {A.}~\bibnamefont {Yacoby}},\ }\bibfield  {title} {\bibinfo {title} {{Imaging
  phonon-mediated hydrodynamic flow in WTe2}},\ }\href {https://doi.org/10.1038/s41567-021-01341-w} {\bibfield  {journal} {\bibinfo  {journal} {Nature Physics}\ }\textbf {\bibinfo {volume} {17}},\ \bibinfo {pages} {1216} (\bibinfo {year} {2021})}\BibitemShut {NoStop}%
\bibitem [{\citenamefont {Valentinis}\ \emph {et~al.}(2023)\citenamefont {Valentinis}, \citenamefont {Baker}, \citenamefont {Bonn},\ and\ \citenamefont {Schmalian}}]{Valentinis2023}%
  \BibitemOpen
  \bibfield  {author} {\bibinfo {author} {\bibfnamefont {D.}~\bibnamefont {Valentinis}}, \bibinfo {author} {\bibfnamefont {G.}~\bibnamefont {Baker}}, \bibinfo {author} {\bibfnamefont {D.~A.}\ \bibnamefont {Bonn}},\ and\ \bibinfo {author} {\bibfnamefont {J.}~\bibnamefont {Schmalian}},\ }\bibfield  {title} {\bibinfo {title} {{Kinetic theory of the non-local electrodynamic response in anisotropic metals: skin effect in 2D systems}},\ }\href {https://doi.org/10.1103/PhysRevResearch.5.013212} {\bibfield  {journal} {\bibinfo  {journal} {Physical Review Research}\ }\textbf {\bibinfo {volume} {5}},\ \bibinfo {pages} {013212} (\bibinfo {year} {2023})}\BibitemShut {NoStop}%
\bibitem [{\citenamefont {Qi}\ and\ \citenamefont {Lucas}(2021)}]{Qi2021}%
  \BibitemOpen
  \bibfield  {author} {\bibinfo {author} {\bibfnamefont {M.}~\bibnamefont {Qi}}\ and\ \bibinfo {author} {\bibfnamefont {A.}~\bibnamefont {Lucas}},\ }\bibfield  {title} {\bibinfo {title} {{Distinguishing viscous, ballistic, and diffusive current flows in anisotropic metals}},\ }\href {https://doi.org/10.1103/PhysRevB.104.195106} {\bibfield  {journal} {\bibinfo  {journal} {Physical Review B}\ }\textbf {\bibinfo {volume} {104}},\ \bibinfo {pages} {195106} (\bibinfo {year} {2021})}\BibitemShut {NoStop}%
\bibitem [{\citenamefont {Varnavides}\ \emph {et~al.}(2022{\natexlab{a}})\citenamefont {Varnavides}, \citenamefont {Wang}, \citenamefont {Moll}, \citenamefont {Anikeeva},\ and\ \citenamefont {Narang}}]{Varnavides2022}%
  \BibitemOpen
  \bibfield  {author} {\bibinfo {author} {\bibfnamefont {G.}~\bibnamefont {Varnavides}}, \bibinfo {author} {\bibfnamefont {Y.}~\bibnamefont {Wang}}, \bibinfo {author} {\bibfnamefont {P.~J.}\ \bibnamefont {Moll}}, \bibinfo {author} {\bibfnamefont {P.}~\bibnamefont {Anikeeva}},\ and\ \bibinfo {author} {\bibfnamefont {P.}~\bibnamefont {Narang}},\ }\bibfield  {title} {\bibinfo {title} {{Mesoscopic finite-size effects of unconventional electron transport in PdCoO$_2$}},\ }\href {https://doi.org/10.1103/PhysRevMaterials.6.045002} {\bibfield  {journal} {\bibinfo  {journal} {Physical Review Materials}\ }\textbf {\bibinfo {volume} {6}},\ \bibinfo {pages} {045002} (\bibinfo {year} {2022}{\natexlab{a}})}\BibitemShut {NoStop}%
\bibitem [{\citenamefont {Varnavides}\ \emph {et~al.}(2022{\natexlab{b}})\citenamefont {Varnavides}, \citenamefont {Jermyn}, \citenamefont {Anikeeva},\ and\ \citenamefont {Narang}}]{Varnavides2022b}%
  \BibitemOpen
  \bibfield  {author} {\bibinfo {author} {\bibfnamefont {G.}~\bibnamefont {Varnavides}}, \bibinfo {author} {\bibfnamefont {A.~S.}\ \bibnamefont {Jermyn}}, \bibinfo {author} {\bibfnamefont {P.}~\bibnamefont {Anikeeva}},\ and\ \bibinfo {author} {\bibfnamefont {P.}~\bibnamefont {Narang}},\ }\bibfield  {title} {\bibinfo {title} {{Probing carrier interactions using electron hydrodynamics}},\ }\Eprint {https://arxiv.org/abs/2204.06004} {arXiv:2204.06004}  (\bibinfo {year} {2022}{\natexlab{b}})\BibitemShut {NoStop}%
\bibitem [{\citenamefont {Ashcroft}\ and\ \citenamefont {Mermin}(1976)}]{Ashcroft1976}%
  \BibitemOpen
  \bibfield  {author} {\bibinfo {author} {\bibfnamefont {N.}~\bibnamefont {Ashcroft}}\ and\ \bibinfo {author} {\bibfnamefont {N.}~\bibnamefont {Mermin}},\ }\href@noop {} {\emph {\bibinfo {title} {{Solid State Physics}}}}\ (\bibinfo  {publisher} {Holt-Saunders},\ \bibinfo {year} {1976})\BibitemShut {NoStop}%
\bibitem [{\citenamefont {Nazaryan}\ and\ \citenamefont {Levitov}(2021)}]{Nazaryan2021}%
  \BibitemOpen
  \bibfield  {author} {\bibinfo {author} {\bibfnamefont {K.~G.}\ \bibnamefont {Nazaryan}}\ and\ \bibinfo {author} {\bibfnamefont {L.}~\bibnamefont {Levitov}},\ }\bibfield  {title} {\bibinfo {title} {{Robustness of vorticity in electron fluids}},\ }\Eprint {https://arxiv.org/abs/2111.09878} {arXiv:2111.09878}  (\bibinfo {year} {2021})\BibitemShut {NoStop}%
\bibitem [{\citenamefont {Reuter}\ and\ \citenamefont {Sondheimer}(1948)}]{Reuter1948}%
  \BibitemOpen
  \bibfield  {author} {\bibinfo {author} {\bibfnamefont {G.~E.~H.}\ \bibnamefont {Reuter}}\ and\ \bibinfo {author} {\bibfnamefont {E.~H.}\ \bibnamefont {Sondheimer}},\ }\bibfield  {title} {\bibinfo {title} {{The theory of the anomalous skin effect in metals}},\ }\href {https://doi.org/10.1098/rspa.1948.0123} {\bibfield  {journal} {\bibinfo  {journal} {Proceedings of the Royal Society A}\ }\textbf {\bibinfo {volume} {195}},\ \bibinfo {pages} {336} (\bibinfo {year} {1948})}\BibitemShut {NoStop}%
\bibitem [{\citenamefont {Dingle}(1953)}]{Dingle1953}%
  \BibitemOpen
  \bibfield  {author} {\bibinfo {author} {\bibfnamefont {R.}~\bibnamefont {Dingle}},\ }\bibfield  {title} {\bibinfo {title} {{The anomalous skin effect and the reflectivity of metals I.}},\ }\href {https://doi.org/10.1016/S0031-8914(54)80042-5} {\bibfield  {journal} {\bibinfo  {journal} {Physica}\ }\textbf {\bibinfo {volume} {19}},\ \bibinfo {pages} {311} (\bibinfo {year} {1953})}\BibitemShut {NoStop}%
\bibitem [{\citenamefont {Lucas}\ and\ \citenamefont {{Das Sarma}}(2018)}]{Lucas2018}%
  \BibitemOpen
  \bibfield  {author} {\bibinfo {author} {\bibfnamefont {A.}~\bibnamefont {Lucas}}\ and\ \bibinfo {author} {\bibfnamefont {S.}~\bibnamefont {{Das Sarma}}},\ }\bibfield  {title} {\bibinfo {title} {{Electronic sound modes and plasmons in hydrodynamic two-dimensional metals}},\ }\href {https://doi.org/10.1103/PhysRevB.97.115449} {\bibfield  {journal} {\bibinfo  {journal} {Physical Review B}\ }\textbf {\bibinfo {volume} {97}},\ \bibinfo {pages} {115449} (\bibinfo {year} {2018})}\BibitemShut {NoStop}%
\bibitem [{\citenamefont {Cook}\ and\ \citenamefont {Lucas}(2019)}]{Cook2019}%
  \BibitemOpen
  \bibfield  {author} {\bibinfo {author} {\bibfnamefont {C.~Q.}\ \bibnamefont {Cook}}\ and\ \bibinfo {author} {\bibfnamefont {A.}~\bibnamefont {Lucas}},\ }\bibfield  {title} {\bibinfo {title} {{Electron hydrodynamics with a polygonal Fermi surface}},\ }\href {https://doi.org/10.1103/PhysRevB.99.235148} {\bibfield  {journal} {\bibinfo  {journal} {Physical Review B}\ }\textbf {\bibinfo {volume} {99}},\ \bibinfo {pages} {235148} (\bibinfo {year} {2019})}\BibitemShut {NoStop}%
\bibitem [{\citenamefont {Ziman}(1960)}]{Ziman1960}%
  \BibitemOpen
  \bibfield  {author} {\bibinfo {author} {\bibfnamefont {J.~M.}\ \bibnamefont {Ziman}},\ }\href@noop {} {\emph {\bibinfo {title} {{Electrons and Phonons: The Theory of Transport Phenomena in Solids}}}}\ (\bibinfo {year} {1960})\BibitemShut {NoStop}%
\bibitem [{\citenamefont {Allen}(1996)}]{Allen1996}%
  \BibitemOpen
  \bibfield  {author} {\bibinfo {author} {\bibfnamefont {P.~B.}\ \bibnamefont {Allen}},\ }\bibfield  {title} {\bibinfo {title} {{Boltzmann Theory and Resistivity of Metals}},\ }in\ \href@noop {} {\emph {\bibinfo {booktitle} {Quantum Theory of Real Materials}}},\ \bibinfo {editor} {edited by\ \bibinfo {editor} {\bibfnamefont {J.~R.}\ \bibnamefont {Chelokowsky}}\ and\ \bibinfo {editor} {\bibfnamefont {S.~G.}\ \bibnamefont {Louie}}}\ (\bibinfo  {publisher} {Springer New York, NY},\ \bibinfo {year} {1996})\ \bibinfo {edition} {1st}\ ed.,\ Chap.~\bibinfo {chapter} {17}, pp.\ \bibinfo {pages} {219--250}\BibitemShut {NoStop}%
\bibitem [{\citenamefont {Guo}\ \emph {et~al.}(2017)\citenamefont {Guo}, \citenamefont {Ilseven}, \citenamefont {Falkovich},\ and\ \citenamefont {Levitov}}]{Guo2017}%
  \BibitemOpen
  \bibfield  {author} {\bibinfo {author} {\bibfnamefont {H.}~\bibnamefont {Guo}}, \bibinfo {author} {\bibfnamefont {E.}~\bibnamefont {Ilseven}}, \bibinfo {author} {\bibfnamefont {G.}~\bibnamefont {Falkovich}},\ and\ \bibinfo {author} {\bibfnamefont {L.~S.}\ \bibnamefont {Levitov}},\ }\bibfield  {title} {\bibinfo {title} {{Higher-than-ballistic conduction of viscous electron flows}},\ }\href {https://doi.org/10.1073/pnas.1612181114} {\bibfield  {journal} {\bibinfo  {journal} {Proceedings of the National Academy of Sciences}\ }\textbf {\bibinfo {volume} {114}},\ \bibinfo {pages} {3068} (\bibinfo {year} {2017})}\BibitemShut {NoStop}%
\bibitem [{\citenamefont {Callaway}(1959)}]{Callaway1959}%
  \BibitemOpen
  \bibfield  {author} {\bibinfo {author} {\bibfnamefont {J.}~\bibnamefont {Callaway}},\ }\bibfield  {title} {\bibinfo {title} {{Model for lattice thermal conductivity at low temperatures}},\ }\href {https://doi.org/10.1103/PhysRev.113.1046} {\bibfield  {journal} {\bibinfo  {journal} {Physical Review}\ }\textbf {\bibinfo {volume} {113}},\ \bibinfo {pages} {1046} (\bibinfo {year} {1959})}\BibitemShut {NoStop}%
\bibitem [{\citenamefont {Scaffidi}\ \emph {et~al.}(2017)\citenamefont {Scaffidi}, \citenamefont {Nandi}, \citenamefont {Schmidt}, \citenamefont {Mackenzie},\ and\ \citenamefont {Moore}}]{Scaffidi2017}%
  \BibitemOpen
  \bibfield  {author} {\bibinfo {author} {\bibfnamefont {T.}~\bibnamefont {Scaffidi}}, \bibinfo {author} {\bibfnamefont {N.}~\bibnamefont {Nandi}}, \bibinfo {author} {\bibfnamefont {B.}~\bibnamefont {Schmidt}}, \bibinfo {author} {\bibfnamefont {A.~P.}\ \bibnamefont {Mackenzie}},\ and\ \bibinfo {author} {\bibfnamefont {J.~E.}\ \bibnamefont {Moore}},\ }\bibfield  {title} {\bibinfo {title} {{Hydrodynamic Electron Flow and Hall Viscosity}},\ }\href {https://doi.org/10.1103/PhysRevLett.118.226601} {\bibfield  {journal} {\bibinfo  {journal} {Physical Review Letters}\ }\textbf {\bibinfo {volume} {118}},\ \bibinfo {pages} {226601} (\bibinfo {year} {2017})}\BibitemShut {NoStop}%
\bibitem [{\citenamefont {Levchenko}\ and\ \citenamefont {Schmalian}(2020)}]{Levchenko2020}%
  \BibitemOpen
  \bibfield  {author} {\bibinfo {author} {\bibfnamefont {A.}~\bibnamefont {Levchenko}}\ and\ \bibinfo {author} {\bibfnamefont {J.}~\bibnamefont {Schmalian}},\ }\bibfield  {title} {\bibinfo {title} {{Transport properties of strongly coupled electron–phonon liquids}},\ }\href {https://doi.org/10.1016/j.aop.2020.168218} {\bibfield  {journal} {\bibinfo  {journal} {Annals of Physics}\ }\textbf {\bibinfo {volume} {419}},\ \bibinfo {pages} {168218} (\bibinfo {year} {2020})}\BibitemShut {NoStop}%
\bibitem [{\citenamefont {Jaggi}(1991)}]{Jaggi1991}%
  \BibitemOpen
  \bibfield  {author} {\bibinfo {author} {\bibfnamefont {R.}~\bibnamefont {Jaggi}},\ }\bibfield  {title} {\bibinfo {title} {{Electron-fluid model for dc size effect}},\ }\href {https://doi.org/10.1063/1.347315} {\bibfield  {journal} {\bibinfo  {journal} {Journal of Applied Physics}\ }\textbf {\bibinfo {volume} {69}},\ \bibinfo {pages} {816} (\bibinfo {year} {1991})}\BibitemShut {NoStop}%
\bibitem [{\citenamefont {Forcella}\ \emph {et~al.}(2014)\citenamefont {Forcella}, \citenamefont {Zaanen}, \citenamefont {Valentinis},\ and\ \citenamefont {van~der Marel}}]{Forcella2014}%
  \BibitemOpen
  \bibfield  {author} {\bibinfo {author} {\bibfnamefont {D.}~\bibnamefont {Forcella}}, \bibinfo {author} {\bibfnamefont {J.}~\bibnamefont {Zaanen}}, \bibinfo {author} {\bibfnamefont {D.}~\bibnamefont {Valentinis}},\ and\ \bibinfo {author} {\bibfnamefont {D.}~\bibnamefont {van~der Marel}},\ }\bibfield  {title} {\bibinfo {title} {{Electromagnetic properties of viscous charged fluids}},\ }\href {https://doi.org/10.1103/PhysRevB.90.035143} {\bibfield  {journal} {\bibinfo  {journal} {Physical Review B}\ }\textbf {\bibinfo {volume} {90}},\ \bibinfo {pages} {035143} (\bibinfo {year} {2014})}\BibitemShut {NoStop}%
\bibitem [{\citenamefont {Beenakker}\ and\ \citenamefont {van Houten}(1988)}]{Beenakker1988}%
  \BibitemOpen
  \bibfield  {author} {\bibinfo {author} {\bibfnamefont {C.~W.}\ \bibnamefont {Beenakker}}\ and\ \bibinfo {author} {\bibfnamefont {H.}~\bibnamefont {van Houten}},\ }\bibfield  {title} {\bibinfo {title} {{Boundary scattering and weak localization of electrons in a magnetic field}},\ }\href {https://doi.org/10.15036/arerugi.37.589} {\bibfield  {journal} {\bibinfo  {journal} {Physical Review B}\ }\textbf {\bibinfo {volume} {38}},\ \bibinfo {pages} {3232} (\bibinfo {year} {1988})}\BibitemShut {NoStop}%
\bibitem [{\citenamefont {Sondheimer}(2001)}]{Sondheimer2001}%
  \BibitemOpen
  \bibfield  {author} {\bibinfo {author} {\bibfnamefont {E.~H.}\ \bibnamefont {Sondheimer}},\ }\bibfield  {title} {\bibinfo {title} {{The mean free path of electrons in metals}},\ }\href {https://doi.org/10.1080/00018730110102187} {\bibfield  {journal} {\bibinfo  {journal} {Advances in Physics}\ }\textbf {\bibinfo {volume} {50}},\ \bibinfo {pages} {499} (\bibinfo {year} {2001})}\BibitemShut {NoStop}%
\bibitem [{\citenamefont {Alekseev}(2016)}]{Alekseev2016}%
  \BibitemOpen
  \bibfield  {author} {\bibinfo {author} {\bibfnamefont {P.~S.}\ \bibnamefont {Alekseev}},\ }\bibfield  {title} {\bibinfo {title} {{Negative Magnetoresistance in Viscous Flow of Two-Dimensional Electrons}},\ }\href {https://doi.org/10.1103/PhysRevLett.117.166601} {\bibfield  {journal} {\bibinfo  {journal} {Physical Review Letters}\ }\textbf {\bibinfo {volume} {117}},\ \bibinfo {pages} {166601} (\bibinfo {year} {2016})}\BibitemShut {NoStop}%
\bibitem [{\citenamefont {Fritz}\ and\ \citenamefont {Scaffidi}(2023)}]{Fritz2023}%
  \BibitemOpen
  \bibfield  {author} {\bibinfo {author} {\bibfnamefont {L.}~\bibnamefont {Fritz}}\ and\ \bibinfo {author} {\bibfnamefont {T.}~\bibnamefont {Scaffidi}},\ }\bibfield  {title} {\bibinfo {title} {{Hydrodynamic electronic transport}},\ }\Eprint {https://arxiv.org/abs/2303.14205} {arXiv:2303.14205}  (\bibinfo {year} {2023})\BibitemShut {NoStop}%
\bibitem [{\citenamefont {Abrikosov}\ and\ \citenamefont {Khalatnikov}(1959)}]{Abrikosov1959}%
  \BibitemOpen
  \bibfield  {author} {\bibinfo {author} {\bibfnamefont {A.~A.}\ \bibnamefont {Abrikosov}}\ and\ \bibinfo {author} {\bibfnamefont {I.~M.}\ \bibnamefont {Khalatnikov}},\ }\bibfield  {title} {\bibinfo {title} {{The theory of a fermi liquid (the properties of liquid 3He at low temperatures)}},\ }\href {https://doi.org/10.1088/0034-4885/22/1/310} {\bibfield  {journal} {\bibinfo  {journal} {Reports on Progress in Physics}\ }\textbf {\bibinfo {volume} {22}},\ \bibinfo {pages} {329} (\bibinfo {year} {1959})}\BibitemShut {NoStop}%
\bibitem [{\citenamefont {Conti}\ and\ \citenamefont {Vignale}(1999)}]{Conti1999}%
  \BibitemOpen
  \bibfield  {author} {\bibinfo {author} {\bibfnamefont {S.}~\bibnamefont {Conti}}\ and\ \bibinfo {author} {\bibfnamefont {G.}~\bibnamefont {Vignale}},\ }\bibfield  {title} {\bibinfo {title} {{Elasticity of an electron liquid}},\ }\href {https://doi.org/10.1103/PhysRevB.60.7966} {\bibfield  {journal} {\bibinfo  {journal} {Physical Review B}\ }\textbf {\bibinfo {volume} {60}},\ \bibinfo {pages} {7966} (\bibinfo {year} {1999})}\BibitemShut {NoStop}%
\bibitem [{\citenamefont {Tokatly}\ and\ \citenamefont {Pankratov}(2000)}]{Tokatly2000}%
  \BibitemOpen
  \bibfield  {author} {\bibinfo {author} {\bibfnamefont {I.}~\bibnamefont {Tokatly}}\ and\ \bibinfo {author} {\bibfnamefont {O.}~\bibnamefont {Pankratov}},\ }\bibfield  {title} {\bibinfo {title} {{Hydrodynamics beyond local equilibrium: Application to electron gas}},\ }\href {https://doi.org/10.1103/PhysRevB.62.2759} {\bibfield  {journal} {\bibinfo  {journal} {Physical Review B}\ }\textbf {\bibinfo {volume} {62}},\ \bibinfo {pages} {2759} (\bibinfo {year} {2000})}\BibitemShut {NoStop}%
\bibitem [{\citenamefont {Valentinis}(2021)}]{Valentinis2021-2}%
  \BibitemOpen
  \bibfield  {author} {\bibinfo {author} {\bibfnamefont {D.}~\bibnamefont {Valentinis}},\ }\bibfield  {title} {\bibinfo {title} {{Optical signatures of shear collective modes in strongly interacting Fermi liquids}},\ }\href {https://doi.org/10.1103/PhysRevResearch.3.023076} {\bibfield  {journal} {\bibinfo  {journal} {Physical Review Research}\ }\textbf {\bibinfo {volume} {3}},\ \bibinfo {pages} {023076} (\bibinfo {year} {2021})}\BibitemShut {NoStop}%
\bibitem [{\citenamefont {Ledwith}\ \emph {et~al.}(2019{\natexlab{a}})\citenamefont {Ledwith}, \citenamefont {Guo}, \citenamefont {Shytov},\ and\ \citenamefont {Levitov}}]{Ledwith2019b}%
  \BibitemOpen
  \bibfield  {author} {\bibinfo {author} {\bibfnamefont {P.}~\bibnamefont {Ledwith}}, \bibinfo {author} {\bibfnamefont {H.}~\bibnamefont {Guo}}, \bibinfo {author} {\bibfnamefont {A.}~\bibnamefont {Shytov}},\ and\ \bibinfo {author} {\bibfnamefont {L.}~\bibnamefont {Levitov}},\ }\bibfield  {title} {\bibinfo {title} {{Tomographic Dynamics and Scale-Dependent Viscosity in 2D Electron Systems}},\ }\href {https://doi.org/10.1103/PhysRevLett.123.116601} {\bibfield  {journal} {\bibinfo  {journal} {Physical Review Letters}\ }\textbf {\bibinfo {volume} {123}},\ \bibinfo {pages} {116601} (\bibinfo {year} {2019}{\natexlab{a}})}\BibitemShut {NoStop}%
\bibitem [{\citenamefont {Pal}\ \emph {et~al.}(2012)\citenamefont {Pal}, \citenamefont {Yudson},\ and\ \citenamefont {Maslov}}]{Pal2012}%
  \BibitemOpen
  \bibfield  {author} {\bibinfo {author} {\bibfnamefont {H.~K.}\ \bibnamefont {Pal}}, \bibinfo {author} {\bibfnamefont {V.~I.}\ \bibnamefont {Yudson}},\ and\ \bibinfo {author} {\bibfnamefont {D.~L.}\ \bibnamefont {Maslov}},\ }\bibfield  {title} {\bibinfo {title} {{Resistivity of non-Galilean-invariant Fermi- and non-Fermi liquids}},\ }\href {https://doi.org/10.3952/physics.v52i2.2358} {\bibfield  {journal} {\bibinfo  {journal} {Lithuanian Journal of Physics}\ }\textbf {\bibinfo {volume} {52}},\ \bibinfo {pages} {142} (\bibinfo {year} {2012})}\BibitemShut {NoStop}%
\bibitem [{\citenamefont {Huang}\ and\ \citenamefont {Lucas}(2021)}]{Huang2021}%
  \BibitemOpen
  \bibfield  {author} {\bibinfo {author} {\bibfnamefont {X.}~\bibnamefont {Huang}}\ and\ \bibinfo {author} {\bibfnamefont {A.}~\bibnamefont {Lucas}},\ }\bibfield  {title} {\bibinfo {title} {{Electron-phonon hydrodynamics}},\ }\href {https://doi.org/10.1103/PhysRevB.103.155128} {\bibfield  {journal} {\bibinfo  {journal} {Physical Review B}\ }\textbf {\bibinfo {volume} {103}},\ \bibinfo {pages} {155128} (\bibinfo {year} {2021})}\BibitemShut {NoStop}%
\bibitem [{\citenamefont {Ledwith}\ \emph {et~al.}(2019{\natexlab{b}})\citenamefont {Ledwith}, \citenamefont {Guo},\ and\ \citenamefont {Levitov}}]{Ledwith2019a}%
  \BibitemOpen
  \bibfield  {author} {\bibinfo {author} {\bibfnamefont {P.~J.}\ \bibnamefont {Ledwith}}, \bibinfo {author} {\bibfnamefont {H.}~\bibnamefont {Guo}},\ and\ \bibinfo {author} {\bibfnamefont {L.}~\bibnamefont {Levitov}},\ }\bibfield  {title} {\bibinfo {title} {{The hierarchy of excitation lifetimes in two-dimensional Fermi gases}},\ }\href {https://doi.org/10.1016/j.aop.2019.167913} {\bibfield  {journal} {\bibinfo  {journal} {Annals of Physics}\ }\textbf {\bibinfo {volume} {411}},\ \bibinfo {pages} {167913} (\bibinfo {year} {2019}{\natexlab{b}})}\BibitemShut {NoStop}%
\bibitem [{\citenamefont {Kawamura}\ and\ \citenamefont {{Das Sarma}}(1992)}]{Kawamura1992}%
  \BibitemOpen
  \bibfield  {author} {\bibinfo {author} {\bibfnamefont {T.}~\bibnamefont {Kawamura}}\ and\ \bibinfo {author} {\bibfnamefont {S.}~\bibnamefont {{Das Sarma}}},\ }\bibfield  {title} {\bibinfo {title} {{Phonon-scattering-limited electron mobilities in Al$_{x}$Ga$_{1-x}$As/GaAs heterojunctions}},\ }\href {https://doi.org/10.1103/PhysRevB.45.3612} {\bibfield  {journal} {\bibinfo  {journal} {Physical Review B}\ }\textbf {\bibinfo {volume} {45}},\ \bibinfo {pages} {3612} (\bibinfo {year} {1992})}\BibitemShut {NoStop}%
\bibitem [{\citenamefont {Ho}\ \emph {et~al.}(2018)\citenamefont {Ho}, \citenamefont {Yudhistira}, \citenamefont {Chakraborty},\ and\ \citenamefont {Adam}}]{Ho2018}%
  \BibitemOpen
  \bibfield  {author} {\bibinfo {author} {\bibfnamefont {D.~Y.}\ \bibnamefont {Ho}}, \bibinfo {author} {\bibfnamefont {I.}~\bibnamefont {Yudhistira}}, \bibinfo {author} {\bibfnamefont {N.}~\bibnamefont {Chakraborty}},\ and\ \bibinfo {author} {\bibfnamefont {S.}~\bibnamefont {Adam}},\ }\bibfield  {title} {\bibinfo {title} {{Theoretical determination of hydrodynamic window in monolayer and bilayer graphene from scattering rates}},\ }\href {https://doi.org/10.1103/PhysRevB.97.121404} {\bibfield  {journal} {\bibinfo  {journal} {Physical Review B}\ }\textbf {\bibinfo {volume} {97}},\ \bibinfo {pages} {121404(R)} (\bibinfo {year} {2018})}\BibitemShut {NoStop}%
\bibitem [{\citenamefont {Aharon-Steinberg}\ \emph {et~al.}(2022)\citenamefont {Aharon-Steinberg}, \citenamefont {V{\"{o}}lkl}, \citenamefont {Kaplan}, \citenamefont {Pariari}, \citenamefont {Roy}, \citenamefont {Holder}, \citenamefont {Wolf}, \citenamefont {Meltzer}, \citenamefont {Myasoedov}, \citenamefont {Huber}, \citenamefont {Yan}, \citenamefont {Falkovich}, \citenamefont {Levitov}, \citenamefont {H{\"{u}}cker},\ and\ \citenamefont {Zeldov}}]{Aharon-Steinberg2022}%
  \BibitemOpen
  \bibfield  {author} {\bibinfo {author} {\bibfnamefont {A.}~\bibnamefont {Aharon-Steinberg}}, \bibinfo {author} {\bibfnamefont {T.}~\bibnamefont {V{\"{o}}lkl}}, \bibinfo {author} {\bibfnamefont {A.}~\bibnamefont {Kaplan}}, \bibinfo {author} {\bibfnamefont {A.~K.}\ \bibnamefont {Pariari}}, \bibinfo {author} {\bibfnamefont {I.}~\bibnamefont {Roy}}, \bibinfo {author} {\bibfnamefont {T.}~\bibnamefont {Holder}}, \bibinfo {author} {\bibfnamefont {Y.}~\bibnamefont {Wolf}}, \bibinfo {author} {\bibfnamefont {A.~Y.}\ \bibnamefont {Meltzer}}, \bibinfo {author} {\bibfnamefont {Y.}~\bibnamefont {Myasoedov}}, \bibinfo {author} {\bibfnamefont {M.~E.}\ \bibnamefont {Huber}}, \bibinfo {author} {\bibfnamefont {B.}~\bibnamefont {Yan}}, \bibinfo {author} {\bibfnamefont {G.}~\bibnamefont {Falkovich}}, \bibinfo {author} {\bibfnamefont {L.~S.}\ \bibnamefont {Levitov}}, \bibinfo {author} {\bibfnamefont {M.}~\bibnamefont {H{\"{u}}cker}},\ and\ \bibinfo {author} {\bibfnamefont {E.}~\bibnamefont {Zeldov}},\ }\bibfield  {title}
  {\bibinfo {title} {{Direct observation of vortices in an electron fluid}},\ }\href {https://doi.org/10.1038/s41586-022-04794-y} {\bibfield  {journal} {\bibinfo  {journal} {Nature}\ }\textbf {\bibinfo {volume} {607}},\ \bibinfo {pages} {74} (\bibinfo {year} {2022})}\BibitemShut {NoStop}%
\bibitem [{\citenamefont {Wolf}\ \emph {et~al.}(2023)\citenamefont {Wolf}, \citenamefont {Aharon-Steinberg}, \citenamefont {Yan},\ and\ \citenamefont {Holder}}]{Wolf2023}%
  \BibitemOpen
  \bibfield  {author} {\bibinfo {author} {\bibfnamefont {Y.}~\bibnamefont {Wolf}}, \bibinfo {author} {\bibfnamefont {A.}~\bibnamefont {Aharon-Steinberg}}, \bibinfo {author} {\bibfnamefont {B.}~\bibnamefont {Yan}},\ and\ \bibinfo {author} {\bibfnamefont {T.}~\bibnamefont {Holder}},\ }\bibfield  {title} {\bibinfo {title} {{Para-Hydrodynamics from weak surface scattering in ultraclean thin flakes}},\ }\href {https://doi.org/10.1038/s41467-023-37966-z} {\bibfield  {journal} {\bibinfo  {journal} {Nature Communications}\ }\textbf {\bibinfo {volume} {14}},\ \bibinfo {pages} {2334} (\bibinfo {year} {2023})}\BibitemShut {NoStop}%
\end{thebibliography}%

\end{document}